\documentclass[lettersize,journal]{IEEEtran}

\def\BibTeX{{\rm B\kern-.05em{\sc i\kern-.025em b}\kern-.08em
    T\kern-.1667em\lower.7ex\hbox{E}\kern-.125emX}}

\usepackage{amsmath,amsfonts}
\usepackage{algorithmicx}
\usepackage{array}
\usepackage[caption=false,font=normalsize,labelfont=sf,textfont=sf]{subfig}
\usepackage{textcomp}
\usepackage{stfloats}
\usepackage{url}
\usepackage{verbatim}
\usepackage{graphicx}
\usepackage{balance}
\usepackage{amsmath,amssymb,amsfonts}
\usepackage{graphicx}
\usepackage{textcomp}
\usepackage{xcolor}
\usepackage{balance}
\usepackage[utf8]{inputenc}
\usepackage{graphicx}
\usepackage{float}
\usepackage{array}
\usepackage{tabularx}
\usepackage{multirow}
\usepackage{amsmath}
\usepackage{amssymb}
\usepackage[flushleft]{threeparttable}
\usepackage{setspace}
\usepackage{color}
\usepackage{cite}
\usepackage{colortbl}
\usepackage{xcolor}
\usepackage{tikzsymbols}
\usepackage{utfsym}
\usepackage{enumitem}
\usepackage{soul}
\usepackage{booktabs}
\usepackage{makecell}
\usepackage{footnote}
\usepackage{algorithm}
\usepackage{algpseudocode}
\usepackage{hyperref}
\usepackage{caption}
\usepackage{diagbox}
\usepackage{amsmath}
\usepackage{tcolorbox}
\usepackage{lipsum}
\usepackage{mdframed}

\newcommand{\insightbox}[1]{
    \begin{tcolorbox}[colback=white, colframe=black, boxrule=0.5pt, arc=2mm, width=0.482\textwidth,
    left=3pt, right=3pt, top=2pt, bottom=2pt]
        #1
    \end{tcolorbox}
}

\begin{document}
\begin{sloppypar}

\title{Explore-Construct-Filter: An Automated Framework for Rich and Reliable API Knowledge Graphs}

\author{Yanbang~Sun,
        Qing~Huang,
        Xiaoxue~Ren,
        Zhenchang~Xing,
        Xiaohong~Li,
        Junjie~Wang

\IEEEcompsocitemizethanks{
\IEEEcompsocthanksitem Y. Sun, X. Li, and J. Wang are with the College of intelligence and computing at Tianjin University, China.
\IEEEcompsocthanksitem Q. Huang, is with the School of Computer Information Engineering at Jiangxi Normal University, China, and is a co-corresponding author along with X. Li.
\IEEEcompsocthanksitem X. Ren is with Zhejiang University, China.
\IEEEcompsocthanksitem Z. Xing is with CSIRO's Data61, Australia.
}
}

\markboth{Journal of \LaTeX\ Class Files,~Vol.~18, No.~9, September~2020}%
{How to Use the IEEEtran \LaTeX \ Templates}

\maketitle

\begin{abstract}
The API Knowledge Graph (API KG) is a structured network that models API entities and their relations, providing essential semantic insights for tasks such as API recommendation, code generation, and API misuse detection.
However, constructing a knowledge-rich and reliable API KG presents several challenges.
Existing schema-based methods rely heavily on manual annotations to design KG schemas, leading to excessive manual overhead.
On the other hand, schema-free methods, due to the lack of schema guidance, are prone to introducing noise, reducing the KG's reliability.
To address these issues, we propose the Explore-Construct-Filter framework, an automated approach for API KG construction based on large language models (LLMs).
This framework consists of three key modules:
1) KG exploration: LLMs simulate the workflow of annotators to automatically design a schema with comprehensive type triples, minimizing human intervention;
2) KG construction: Guided by the schema, LLMs extract instance triples to construct a rich yet unreliable API KG;
3) KG filtering: Removing invalid type triples and suspicious instance triples to construct a rich and reliable API KG.
Experimental results demonstrate that our method surpasses the state-of-the-art method, achieving a 25.2\% improvement in F1 score.
Moreover, the Explore-Construct-Filter framework proves effective, with the KG exploration module increasing KG richness by 133.6\% and the KG filtering module improving reliability by 26.6\%.
Finally, cross-model experiments confirm the generalizability of our framework.
\end{abstract}


\section{INTRODUCTION}
Application Programming Interfaces (APIs) are essential in modern software development, enabling seamless interactions between system components \cite{raatikainen2021state, wu2024future, lercher2024microservice}.
Even simple programs, such as ``Hello World,'' rely on at least one API.
However, the vast number of available APIs and their complex interconnections pose significant challenges for developers.
For example, the Java standard library contains over 30,000 APIs, and any two APIs may exhibit as many as 11 distinct types of semantic relations \cite{javaapi, huang2022se}.
In such a complex ecosystem, developers often struggle to choose and use the right APIs.
Take \textit{HashMap} and \textit{Hashtable} as an example:
both store key-value pairs, but \textit{HashMap} is non-thread-safe, while \textit{Hashtable} is thread-safe.
Ignoring these subtle differences can lead developers to choose the wrong API, resulting in inefficient or incorrect implementations.
Therefore, organizing the interconnections between APIs (such as similar usage and different behaviors) into structured API knowledge will enhance developers' ability to make informed API choices.

Previous studies \cite{Manual, Huang2018APIMR, Liu2020GeneratingCB} introduce the concept of an API Knowledge Graph (KG), which organizes and represents the rich semantic information among APIs.
API KGs offer a valuable approach for understanding and leveraging API knowledge in various scenarios. 
For example, in API recommendation tasks~\cite{Liu2023RecommendingAA, Peng2021RevisitingBA, Huang2018APIMR}, when the order of set elements is not required, an API KG can recommend the more efficient \textit{java.util.HashSet} over \textit{java.util.TreeSet}.
This is because the KG not only captures the usage of APIs but also reveals subtle differences between them, enabling more informed and context-aware recommendations.
However, despite their great potential, constructing a knowledge-rich and reliable API KG still faces many challenges

Existing methods for constructing API KGs are predominantly schema-based~\cite{huang2022se, yanbang1, Ren2020APIMisuseDD}.
These methods rely on predefined schemas, including entity types, relation types, and type triples (combinations of entity types and relation types), to guide the extraction of API entities and relations.
For example, Huang et al.~\cite{yanbang1} design a schema that includes 3 entity types, 9 relation types, and 9 type triples by annotating and summarizing the entity types and relation types in API and tutorial documentation, and construct an API KG based on this schema.
However, developing such schemas requires annotators to invest significant time, resulting in high labor costs.
In particular, the more relation types are needed for constructing a rich KG, the more annotated documentation there will be, which further exacerbates the labor costs~\cite{yanbang2}.

In the field of natural language processing, schema-free methods \cite{GraphRAG, EDC, free3, free4} are another mainstream approach for constructing KGs.
Unlike schema-based methods, schema-free methods extract instance triples directly from text, thereby reducing labor costs.
For example, Zhang et al.~\cite{EDC} propose EDC, which utilizes large language models (LLMs) through few-shot prompting to identify and extract instance triples ([subject, relation, object]) from input texts, independent of any specific schema.
However, when these methods are applied to the construction of API KGs, they often introduce noise and reduce the reliability of the KG.
Due to the lack of schema guidance, these methods may extract inaccurate instance triples.
For example, the instance triple (\textit{SortedSet, is a type of, Set}) extracted from the text ``You can add elements to a SortedSet just like you would to a regular Set...'' fails to accurately express the original semantic meaning.
Furthermore, the constructed KG lacks type information, which limits its practicality.
For example, it is impossible to retrieve classes that are similar in usage to the class \textit{SortedSet}.

In summary, both schema-based and schema-free methods face significant challenges in constructing API KGs.
These challenges primarily stem from two aspects: schema-based methods suffer from the high labor costs associated with designing schemas, while schema-free methods struggle due to the absence of schemas during the extraction of API entities and relations.
To address these issues, it is crucial to automate the schema design and construct the API KG based on the generated schema.
Therefore, we propose using large language models (LLMs) to automate the construction of API KGs.
First, LLMs simulate the annotator's workflow in a bottom-up manner, starting by labeling specific types and then abstracting them into high-level types, thus constructing the schema.
Then, based on this schema, the LLM can extract API entities and relations to construct the API KG.
Through pre-training, LLMs store extensive API knowledge and demonstrate excellent capabilities in information extraction~\cite{yanbang1, yanbang2}, making them a promising solution for overcoming these challenges.

Despite the powerful potential of LLMs, they still face challenges when handling multi-step complex reasoning tasks, such as API KG construction.
Especially for KG schema design, which involves several interrelated tasks like entity identification, entity type labeling, and entity type fusion, LLMs struggle to complete these tasks all at once.
To address this, we adopt a Chain-of-Thought (CoT~\cite{cot1, cot2, cot3, yanbang2}) strategy, breaking the KG construction task into simple tasks and sequentially calling the LLM to complete them.
This step-by-step reasoning strategy enhances the LLM's reasoning accuracy, ensuring the construction of a more reliable KG.

However, due to the ``hallucination'' issue of LLMs~\cite{huang2024survey, dhuliawala2023chain}, their performance may not be reliable for data processing tasks that require high precision, such as collecting all identified relations.
To address this, we introduce a hybrid execution strategy.
Specifically, we implement non-AI units through coding to handle tasks that require high precision, while the AI units, driven by LLMs, focus on reasoning tasks.
These units are logically organized into an AI chain, ensuring that each task in the KG construction is effectively performed.

To ensure the comprehensiveness of the KG, we adopt a fully connected strategy to construct a KG schema with diverse type triples.
Specifically, all entity types and relation types are combined to generate as many potential type triples as possible.
For example, entity types $ET_1$, $ET_2$, and relation type $RT_1$ can be combined into two type triples: ($ET_1$, $RT_1$, $ET_2$) and ($ET_2$, $RT_1$, $ET_1$).
However, due to the absence of manual verification, the KG schema generated by LLMs may not fully align with expert judgment, potentially introducing invalid type triples that result in suspicious instance triples.
For example, the type triple (\textit{class}, \textit{dependency}, \textit{method}) is invalid because, class dependencies typically point to other classes or interfaces, not specific methods~\cite{martin2000design}.
This leads to suspicious instance triples like (\textit{FileWriter}, \textit{relies on}, \textit{flush()}).
Therefore, we apply association rules to the construction of the KG.
By calculating the association strength (using \textit{Support}, \textit{Confidence}, and \textit{Lift} as measurement metrics) between entity types and relation types within the type triples, we can remove invalid triples that fall below the predefined thresholds for each metric, thereby filtering out suspicious instance triples to ensure the reliability of the API KG.

In conclusion, our framework consists of three key modules: KG exploration, KG construction, and KG filtering.
Specifically, the KG exploration module generates potential KG schemas, containing comprehensive entity types, relation types, and type triples.
The KG construction module extracts API entities and relations based on the schema, constructing a rich but unreliable KG.
Finally, the KG filtering module removes invalid type triples and suspicious instance triples, resulting in a rich and reliable  KG.
Among these, both the KG exploration and KG construction modules contain both AI and non-AI units.
For example, the KG exploration module includes the AI unit for entity type labeling and the non-AI unit for fully connected KG schema generation.
To ensure the reliability of the filtered KG, the KG filtering module is composed entirely of non-AI units, including the KG schema update unit and the KG update unit.

We systematically conduct experiments to evaluate our method’s performance.
First, we determine the most balanced thresholds for the KG filtering module, with \textit{Support}, \textit{Confidence}, and \textit{Lift} set to 0.005, 0.02, and 1.0, respectively. 
This configuration achieves a type triple accuracy of 0.76 and retains 26 valid type triples.
Second, our method significantly outperforms the state-of-the-art EDC~\cite{EDC}, achieving an F1 improvement of 25.2\% for API KG construction.
Moreover, the explore-construct-filter framework proves effective: the KG exploration module enhances the richness of KG by 133.6\%, the KG filtering module improves the reliability of the KG by 26.6\%, and the fully connected strategy increases the comprehensiveness of the KG by 33.5\%.
Finally, experiments across different LLMs (GPT-4, Llama, and Claude) demonstrate that our method exhibits strong generalizability.

In this paper, we make the following contributions:
\begin{itemize}[leftmargin=*]
    \item 
    Conceptually, we propose a method for automatically constructing an API KG.
    This method simulates the workflow of manual annotators using LLMs, automatically designs the KG schema, and then constructs the API KG. This method reduces the manual effort required to design the schema and enhances the efficiency of KG construction.
    \item
    Methodologically, we design the ``Explore-Construct-Filter'' framework to achieve this automation goal.
    The framework leverages LLMs to discover diverse entity types, relation types, and type triples, then extracts instance triples to construct a comprehensive KG.
    Through a filtering mechanism, the framework removes suspicious instance triples, ensuring that the constructed KG is both comprehensive and reliable.
    \item
    Technologically, we introduce a CoT strategy, which improves LLM's accuracy through step-by-step reasoning.
    Moreover, we adopt a hybrid execution strategy to ensure that each task is performed efficiently.
\end{itemize}

\section{Background}\label{sec: motivation}
In this section, we first introduce API relations and API KG construction, and then discuss how to use LLMs to enhance the API KG construction process.

\subsection{API Relation}
API relations reflect the semantic connections and interactions between different APIs.
Liu et al.~\cite{Liu2020GeneratingCB} studied the subtle differences between APIs, comparing them from three aspects: categorization, functionality, and characteristics.
Based on this, Huang et al.~\cite{huang2022se} summarized nine types of API semantic relations, including behavior-difference, function-replace, and efficiency-comparison.
For example, the behavior difference relation describes how two similar APIs behave differently when performing the same task, whereas the efficiency comparison relation specifies the variance in efficiency between two APIs under particular conditions.
These API semantic relations are widely present in natural language texts (e.g., API documentation and Q\&A forums) and understanding them is crucial for the correct and efficient use of APIs.
In this paper, we focus more on the semantic relations between APIs rather than their structural relations.
The former requires careful mining from large amounts of textual data, while the latter can be easily obtained from development documentation.

\subsection{API KG Construction}\label{sec: kgcon}
The API KG is a complex network structure designed to represent APIs and their interrelationships.
In an API KG, each node corresponds to an API and includes essential attributes such as name, description, and functionality.
This information enables developers to quickly understand the API's purpose and usage.
The edges in the graph represent semantic relations between APIs, such as constraint and collaboration.
By analyzing these relations, developers can gain a clear understanding of how APIs interact, optimizing system integration.

API KG construction begins with the design of the KG schema, which defines the entity and relation types within the KG.
This schema uses type triples, such as (\textit{$ET_{A}$, $RT_{R}$, $ET_{B}$}), to represent connections between entity types A and B through relation type R.
Guided by this schema, instances are extracted from diverse data sources to form the API KG, represented by instance triples like (\textit{$e_{a}$, $r$, $e_{b}$}), where entity \textit{a} is related to entity \textit{b} via relation \textit{r}.
Therefore, an API KG with a rich set of instances requires a complete KG schema that includes as many type triples as possible.

Most API KG schemas are manually designed by annotators~\cite{huang2022se,Huang2018APIMR,Li2018ImprovingAC}.
These annotators determine the necessary entity and relation types based on domain knowledge and summarize type triples to construct the schema.
However, developing these schemas requires significant time from annotators, resulting in high labor costs.
In this paper, we propose an LLM-based automated method that simulates annotators in summarizing entity and relation types from a small set of texts to generate a complete schema.
Based on it, we further construct a knowledge-rich API KG.

\begin{figure*}[t]
    \centering
    \includegraphics[width=1.0\linewidth]{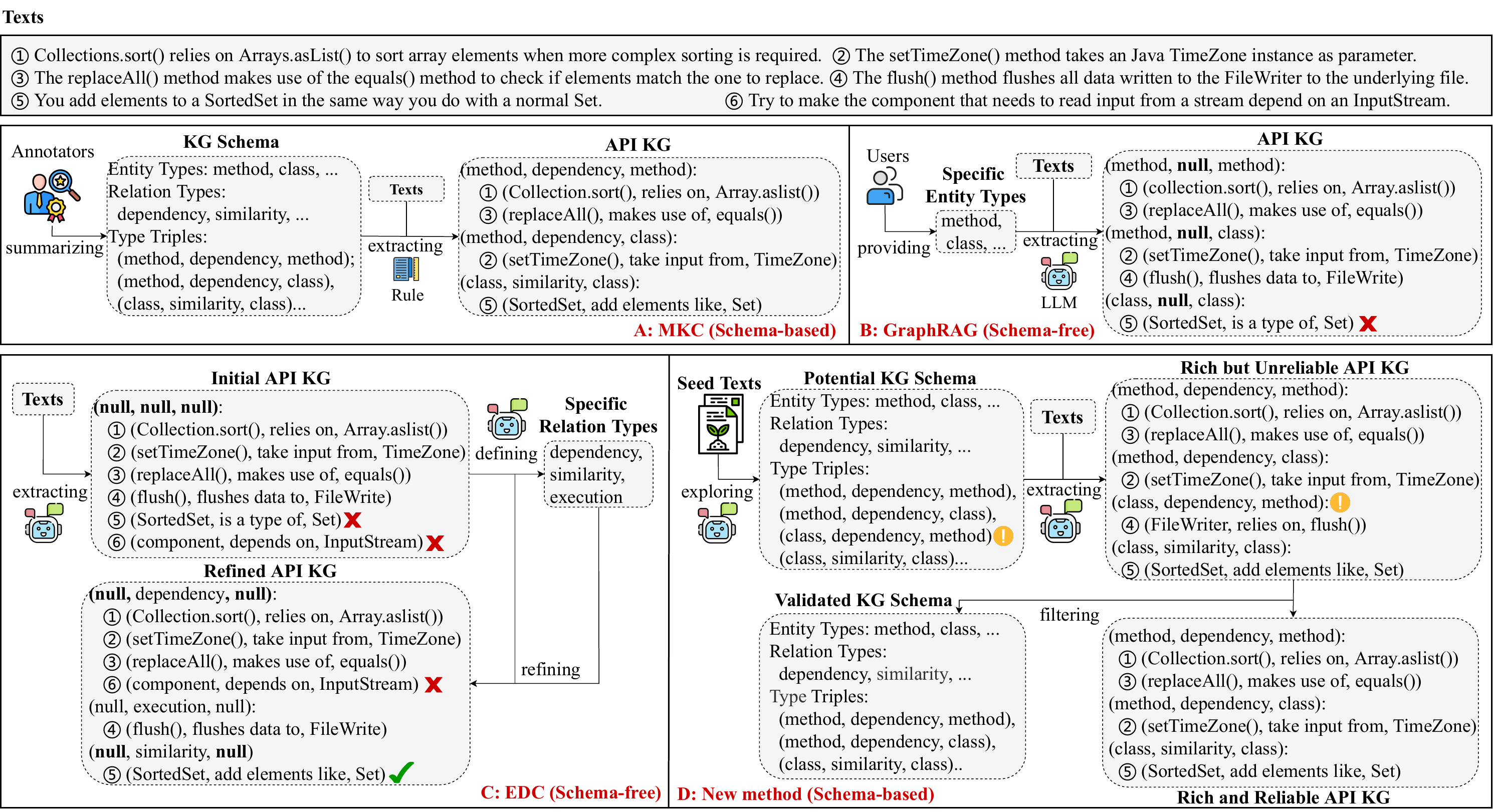}    
    \caption{The Comparison of API KG Construction Methods.}
    \label{fig: killing_example}
\end{figure*}

\subsection{LLM for KG Construction}
LLMs, such as GPT-4~\cite{GPT4oAnalysis-1} and Claude~\cite{claude-3-5-sonnet}, are deep learning models trained on vast amounts of text data.
Recently, LLMs are widely applied to downstream tasks.
For instance, some researchers~\cite{yanbang1, yanbang2} attempt to utilize LLMs to extract API entities and relations, thereby constructing an API KG.
However, no work has yet attempted to use LLMs to design a KG schema.
In the schema design phase, LLMs need to reason about various components of the KG schema, such as how to categorize entities and how to define relations between them.
To improve reasoning accuracy, we use the Chain-of-Thought (CoT) method, which breaks down complex tasks into multiple simple tasks, and leverages the LLM to accomplish these simple tasks step by step.

To further enhance the effectiveness of task execution, we introduce the in-context learning method~\cite{Brown2020LanguageMA, Min2022RethinkingTR}.
By providing task descriptions and examples, this method helps LLMs capture patterns and rules within tasks.
However, research~\cite{PCR, huang2024se} shows that the effectiveness of in-context learning largely depends on the design of prompts, including prompt style, example content, and example order.
To address it, we adopt structured prompts~\cite{xing2025when} to enhance the LLM's performance across various tasks.
By combining these strategies, LLMs can better simulate manual annotators to construct API KGs.

\section{Motivation}
In this section, we will provide a detailed explanation of the limitations of existing methods.
As shown in Fig.~\ref{fig: killing_example}, six texts are randomly selected from Stack Overflow posts, each describing the semantic relations between Java APIs.
Both methods are applied to construct KGs from these texts.

Fig.~\ref{fig: killing_example}-A illustrates the workflow of the representative schema-based method, short for MKC~\cite{huang2022se}.
First, this method relies on annotators to summarize entity types, relation types, and type triples to design the KG schema.
Based on this schema, preset rules (e.g., API\_1 be similar to API\_2) are then used to extract instance triples from the texts.
Finally, the method constructs an API KG containing instance triples with their type information.
For example, the entity type of \textit{Collection.sort()} is method, and the relation type of ``relies on'' is dependency.
Note that no instance triples are extracted from the fourth and sixth texts, as they do not match any type triples in the schema.
However, this method heavily depends on annotators, leading to high labor costs.

One representative schema-free method is GraphRAG~\cite{GraphRAG}.
As shown in Fig.~\ref{fig: killing_example}-B, GraphRAG requires users to specify the entity type according to the task.
Based on these entity types, an LLM is used to extract instance triples from the text, constructing a KG containing five instance triples.
However, due to the lack of guidance on relation types, LLMs fail to accurately analyze the relation between entities.
Therefore, the extracted instance triples are inconsistent with the text semantics, introducing noise and thus reducing the reliability of the KG.
For example, this method extracts the incorrect instance triple (\textit{SortedSet, is a type of, Set}) from the fifth text.
Furthermore, these instance triples lack relation types, which limits the utility of the KG.
For example, the instance triple (\textit{collection.sort(), relies on, Array.aslist()}) belongs to the type triple (\textit{method, null, method}).

Zhang et al.~\cite{EDC} propose another schema-free method, EDC, with its workflow illustrated in Fig.~\ref{fig: killing_example}-C.
First, the method uses an LLM to extract instance triples, forming an initial KG.
It then clusters the extracted relations and defines relation types (e.g., dependency and similarity).
Based on these relation types, EDC refines the initial KG.
By using a text similarity-based search, it recommends candidate relation types to help correct errors made during the extraction phase.
For example, (\textit{SortedSet, is a type of, Set}) can be corrected to (\textit{SortedSet, add elements like, Set}) due to the candidate relation type ``similarity''.
However, although refining phase can correct extracted relations, it cannot correct extracted entities due to the lack of entity types.
Therefore, there is still noise in the KG, which reduces the reliability of the KG.
For example, (\textit{component, depends on, InputStream}) is incorrect because component is not considered an API entity.
Furthermore, the lack of entity types also limits the utility of the KG.

To overcome these limitations, we propose a novel schema-based method, as shown in Fig.~\ref{fig: killing_example}-D.
To reduce labor costs, we use LLMs to extract instance triples from a small amount of user-provided seed text and explore the entity and relation types to generate a potential KG schema.
Based on this schema, LLMs extract instance triples from the target text (e.g., the given six texts) to construct a KG with complete type information.
This KG contains five instance triples (the sixth text does not match any type triples), each labeled with its type information.
However, since our method lacks manual verification, the generated schema may contain some invalid type triples, such as the type triple (\textit{class}, \textit{dependency}, \textit{method}).
This is because, in object-oriented design, class dependencies are typically directed towards other classes or interfaces, rather than specific methods.
Consequently, the KG constructed based on this schema contains suspicious instance triples, such as (\textit{FileWriter}, \textit{relies on}, \textit{flush()}).  
Therefore, it is necessary to verify the potential schema.

Therefore, we apply association rules to the KG construction by calculating the association strength of each type triple and filtering out those with association strength below a certain threshold, along with their instance triples.
Ultimately, we obtain a validated KG schema (including 3 type triples) and a reliable API KG (including 4 instance triples).
Our method can automate the design of the KG schema, thereby alleviating the issue of high labor costs.
Meanwhile, the schema-based approach minimizes noise as much as possible and ensures that the KG contains complete type information.

\section{APPROACH}

\begin{figure*}[t]
    \centering
    \includegraphics[width=1.0\linewidth]{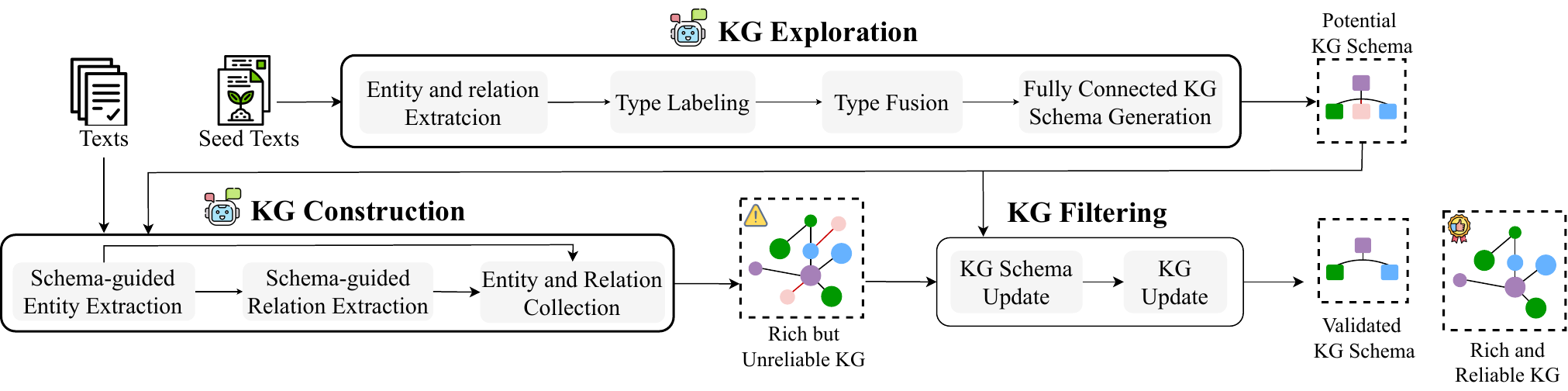}
    \caption{Overall Framework of Our Method.}
    \label{fig: approach}
\end{figure*}

In this paper, we propose an LLM-based automated method for API KG construction.
The overall framework of our method is shown in Fig.~\ref{fig: approach}, which consists of three modules: KG exploration, KG construction, and KG filtering.
The KG exploration module thoroughly analyzes entity types and relation types from seed texts.
It then combines entity types and relation types comprehensively using a fully connected strategy to form a KG schema containing all potential type triples.
The KG construction module extracts instance triples based on this schema to construct a rich but unreliable KG, which may contain some suspicious instance triples.
The KG filtering module then removes these suspicious instances using frequency-based statistics, constructing a rich and reliable KG.
Next, we will introduce each module in detail.


\subsection{KG Exploration}
\begin{figure}[t]
    \centering
    \includegraphics[width=1.0\linewidth]{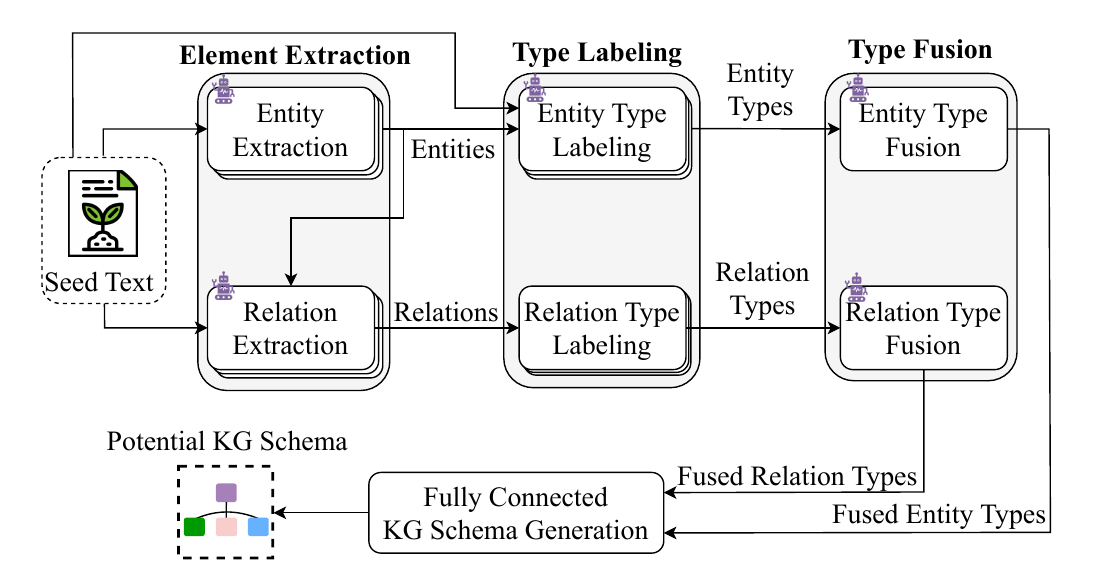}
    \caption{Workflow of KG Exploration Module.}
    \label{fig: module1}
\end{figure}

To construct a richness API KG, the primary task is to design a KG schema that contains diverse entity types and relation types.
To achieve this, we design the KG exploration module, which follows the principle of ``abstracting high-dimensional types from low-dimensional facts'' to automatically design the KG schema in a bottom-up manner.
As shown in Fig.~\ref{fig: module1}, the KG Exploration module consists of seven functional units: entity extraction, relation extraction, entity type labeling, relation type labeling, entity type fusion, relation type fusion, and fully connected KG schema generation.
Among them, except for the relation type labeling and the fully connected KG schema generation unit, the other units are all AI units, which implement specific functions by calling LLMs.  
The input of the KG exploration module is a set of seed texts, and the output is a potential KG schema.
Next, we will provide a detailed description of the functional units of this module to help understand its workflow.

\subsubsection{Entity Extraction}
This unit is used to extract API entity instances of any type from the given text.
As shown in Fig.~\ref{fig: module1}, the input of this unit is natural language text, and the output is the API entities contained in the text.
To improve the performance of LLMs, based on the extensive prompting pattern proposed by Xing et al.~\cite{xing2025when}, we format the natural language prompt into a controllable structured prompt. 
For each AI unit of our framework, we provide a detailed description of its prompt design, including role descriptions, task commands, and considerations, as well as four illustrative examples.
For more details, please refer to the Appendix~\ref{sec: appendix}.
The structured prompt design for this unit is shown in Fig.~\ref{fig: ee}.

\subsubsection{Relation Extraction}\label{sec: relation extraction}
\begin{figure}[h]
    \centering
    \includegraphics[width=1.0\linewidth]{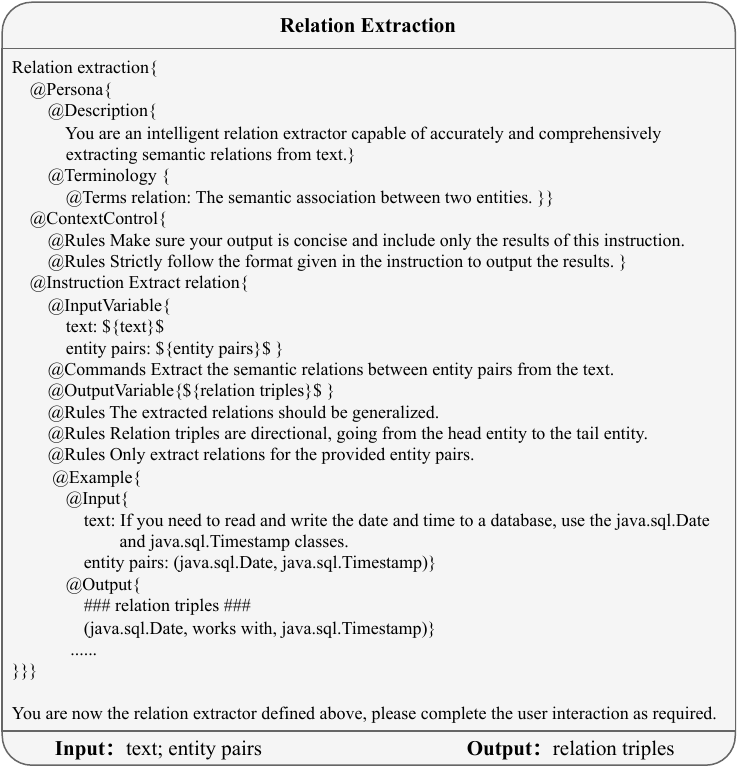}
    \caption{Prompt for Relation Extraction Unit.}
    \label{fig: re}
\end{figure}

To extract the relations between API entities more accurately, similar to previous works~\cite{yanbang2}, we combine the extracted entities into API pairs. 
For example, \textit{$e_{1}$}, \textit{$e_{2}$}, and \textit{$e_{3}$} can be combined into (\textit{$e_{1}$}, \textit{$e_{2}$}), (\textit{$e_{2}$}, \textit{$e_{3}$}), and (\textit{$e_{1}$}, \textit{$e_{3}$}).
As shown in Fig.~\ref{fig: module1}, we input the text and API entity pairs into the relation extraction unit at the same time, it outputs the instance triples in the form of (\textit{$e_{head}$}, \textit{r}, \textit{$e_{tail}$}).
The structured prompt design for this unit is shown in Fig.~\ref{fig: re}.

\subsubsection{Entity Type Labeling}
\begin{figure}[h]
    \centering
    \includegraphics[width=1.0\linewidth]{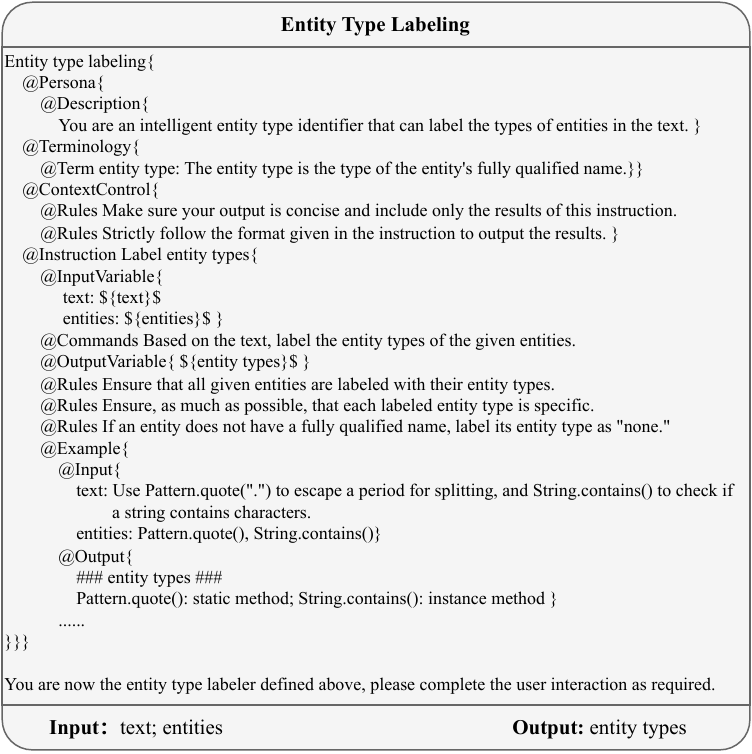}
    \caption{Prompt for Entity Type Labeling Unit.}
    \label{fig: etl}
\end{figure}

This unit is used to label the entity types of API entities. 
As shown in Fig.~\ref{fig: module1}, its input is the text and the extracted API entities, and its output is the entity type to which each entity belongs.
In order to achieve the goal of abstracting high-dimensional types from low-dimensional facts, this unit should output specific and low-dimensional entity types, such as concrete class, utility class, etc.
Please refer to Fig.~\ref{fig: etl} for more detailed information.

\subsubsection{Relation Type Labeling}\label{sec: Relation Type Labeling}
The relation type labeling unit is used to label the relation types of the relation. 
Since the relation instances in the instance triples are concise enough, they can be regarded as low-dimensional relation types.
In short, the relation types here are exactly the relation instances in the instance triples, so this unit is a non-AI unit and does not require the participation of the LLM.

\subsubsection{Entity Type Fusion}
\begin{figure}[htbp]
    \centering
    \includegraphics[width=0.95\linewidth]{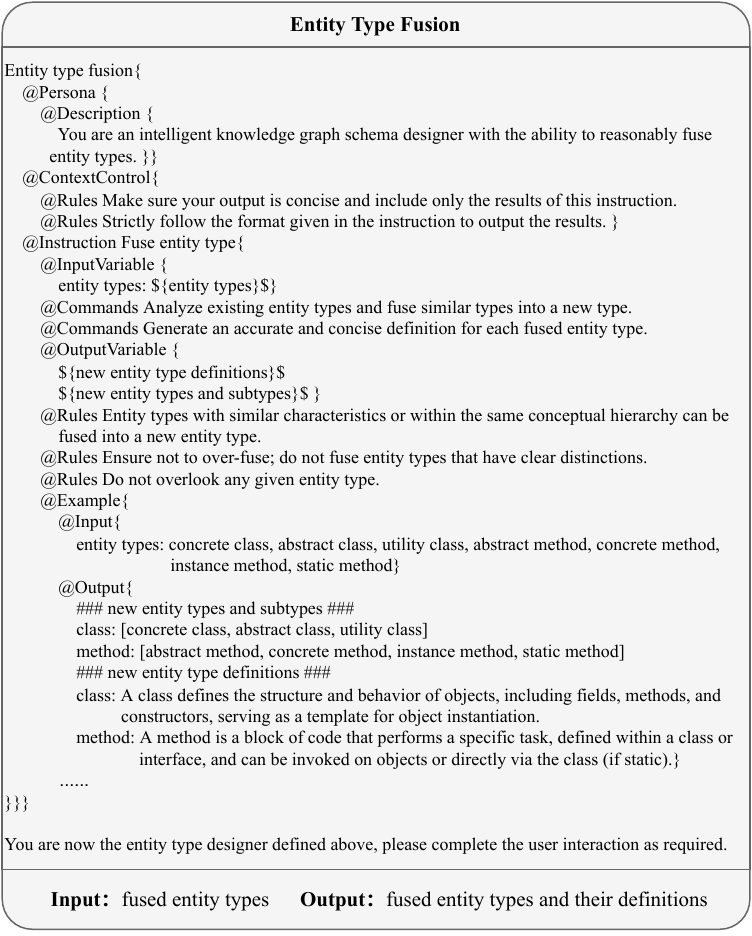}
    \caption{Prompt for Entity Type Fusion Unit.}
    \label{fig: etf}
\end{figure}

As shown in Fig.~\ref{fig: module1}, this unit can only start after the entity types in all the seed texts have been labeled.
It takes all low-dimensional entity types as input and outputs new high-dimensional entity types.
For example, concrete class and utility class are fused into the class category.
To improve the accuracy of subsequent schema-guided entity extraction, this unit also generates definitions for each fused entity type.
Moreover, it outputs the mapping between the new entity types and the original types, such as ``class: [concrete class, utility class]'', for performance evaluation.
The prompt design for this unit can be seen in Fig.~\ref{fig: etf}.

\subsubsection{Relation Type Fusion}
\begin{figure}[htbp]
    \centering
    \includegraphics[width=0.95\linewidth]{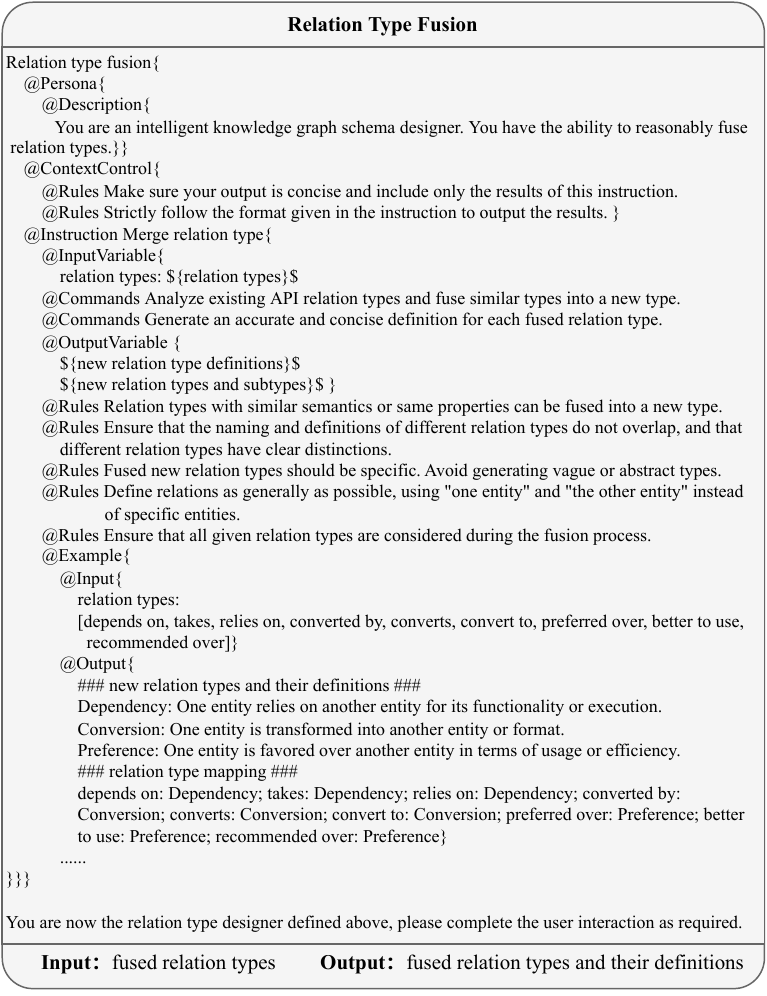}
    \caption{Prompt for Relation Type Fusion Unit.}
    \label{fig: rtf}
\end{figure}

The relation type fusion unit abstracts low-dimensional relation types into high-dimensional ones. 
It takes all low-dimensional relation types as input and outputs new high-dimensional relation types.
For example, ``relies on'' and ``depends on'' can be fused into new relation type ``dependency''
This unit can only start after all relation types have been extracted from the seed texts.
It also outputs the definition of the new relation type and the mapping between the new relation type and the low-dimensional relation types.
More details of this unit can be found in Fig.~\ref{fig: rtf}.

\subsubsection{Fully Connected KG Schema Generation}
This unit is used to construct the KG schema, which guides the subsequent schema-based entity extraction and relation extraction units.
Its input consists of the fused entity types and relation types, and its output is a KG schema containing type triples.
In order to construct an API KG with rich API knowledge, we would like to mine as many type triples as possible.
Therefore, this unit adopts a full combination strategy, combining all entity types and relation types to generate all potential type triples.
For example, given two fused entity types \textit{$ET_1$} and \textit{$ET_2$}, and two fused relation types \textit{$RT_1$} and \textit{$RT_2$}, they can be combined into four potential type triples, such as (\textit{$ET_1$}, \textit{$RT_1$}, \textit{$ET_2$}).

\vspace{-2mm}
\subsection{KG Construction}
\begin{figure}[t]
    \centering
\includegraphics[width=1.0\linewidth]{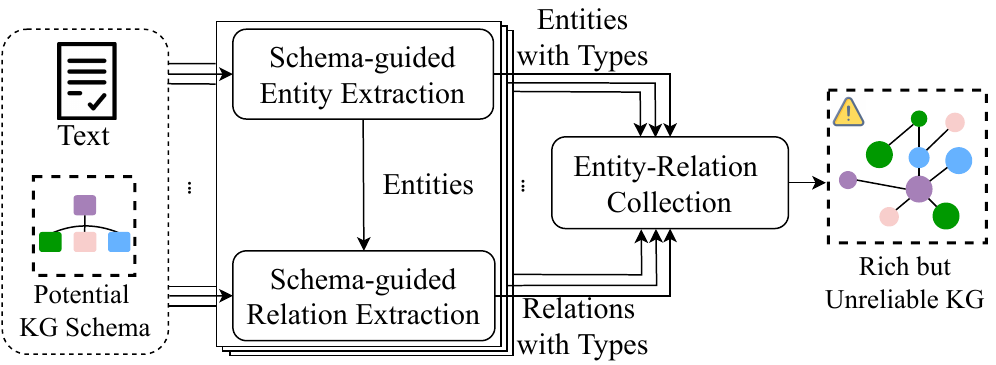}
    \caption{Workflow of KG Construction Module.}
    \label{fig: module2}
\end{figure}

To construct API KG based on KG schema, we design the KG construction module.
As shown in Fig.~\ref{fig: module2}, this module consists of three functional units: schema-guided entity extraction, schema-guided relation extraction, and entity-relation collection. 
The first two are AI units while entity-relation collection is a non-AI unit.
Next, we will describe these units.

\subsubsection{Schema-guided Entity Extraction}
\begin{figure}[H]
    \centering
    \includegraphics[width=0.95\linewidth]{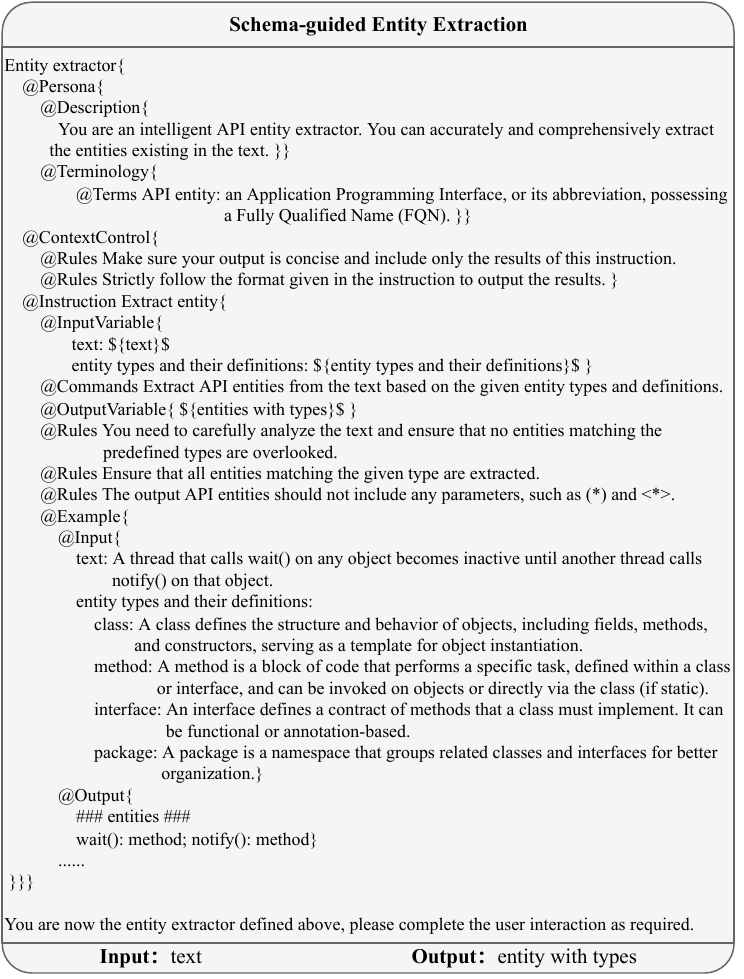}
    \caption{Prompt for Schema-guided Entity Extraction Unit.}
    \label{fig: see}
\end{figure}

This unit is used to extract API entities of a given type from the text. 
Its input is the text, entity types and their definitions, and its output is API entity instances and the entity types to which they belong.
The prompt design of this unit can be seen in Fig.~\ref{fig: see}.

\subsubsection{Schema-guided Relation Extraction}
\begin{figure}[H]
    \centering
    \includegraphics[width=0.9\linewidth]{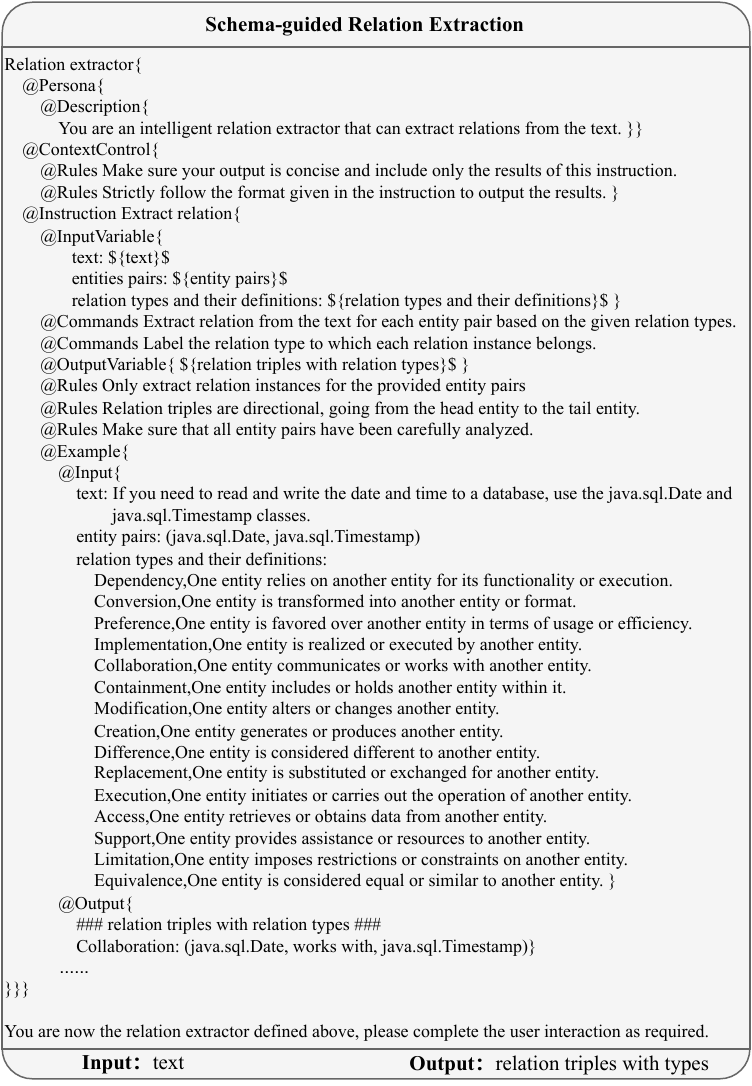}
    \caption{Prompt for Schema-guided Relation Extraction Unit.}
    \label{fig: sre}
\end{figure}

To improve the accuracy of relation extraction, we combine the extracted API entities into API pairs.
Then, we input the text, API entity pairs, and the relation types into this unit, which output are the extracted API relations and the relation types to which they belong.
Please refer to Fig.~\ref{fig: sre} for more details.

\subsubsection{Entity-Relation Collection}
After extracting entities and relations from all the texts, this unit will collect all the instance triples and their type information to construct the API KG.

\vspace{-2mm}
\subsection{KG Filtering}
\begin{figure}[t]
    \centering
\includegraphics[width=1.0\linewidth]{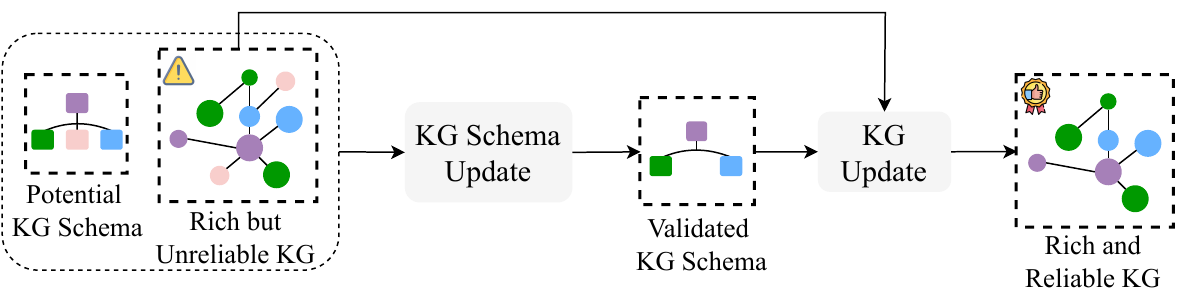}
    \caption{Workflow of KG Filtering Module.}
    \label{fig: module3}
    \vspace{-4mm}
\end{figure}

Since the KG exploration module is fully automated and lacks manual verification of the KG schema, there may be many invalid type triples.
As a result, the KG constructed based on this schema may contain suspicious instance triples.
To remove these suspicious instance triples, we design a KG filtering module.
As shown in Fig.~\ref{fig: module3}, this module consists of two non-AI units: KG schema update and KG update.

\subsubsection{KG Schema Update}
This unit is used to remove invalid type triples in the KG schema.
Its inputs are the unreliable KG and the potential KG schema, and its output is the validated KG schema.
In this module, we adopt a frequency-based method to access the validity of type triples.

In the field of data mining, the association rule is often used to measure the association strength between different items~\cite{asso1, asso2, asso3}.
Inspired by this, we apply the association rule to evaluate the association strength between entity types and relation types in type triples, that is, the validity of type triples.
However, there are various ways to construct association rules. 
For example, the relation type can be inferred from the entity type pair, that is, ($ET_{1}$, $ET_{2}$) $\rightarrow$ $RT_{1}$;
another entity type can also be inferred from a certain entity type and the relation type, that is, ($ET_{1}$, $RT_{1}$) $\rightarrow$ $ET_{2}$.
Since this paper mainly focuses on the potential relation types between entity types, the \textit{Pattern} ($ET_{1}$, $ET_{2}$) $\rightarrow$ $RT_{1}$ is adopted.
For this \textit{Pattern}, the method includes three metrics:

\begin{itemize}[leftmargin=*]
    \item \textit{Support}: It refers to the proportion of the number of type triples 
    $\mathit{(ET_{1}, RT_{1}, ET_{2})}$ in the KG, which can be used to measure the universality of the type triple.
    Its calculation formula is as follows, where \textit{all} refers to the total number of type triples.
    \[
    \small \textit{Support}\left( \textit{Pattern} \right) = \frac{\textit{num}\left( \mathit{ET_1, RT_1, ET_2} \right)}{\textit{all}}
    \]
    \item \textit{Confidence}: It refers to the conditional probability of the appearance of the relation type $RT_{1}$ under the condition that the entity types $ET_{1}$ and $ET_{2}$ already exist, which can be used to measure the reliability of the type triple. 
    Its calculation formula is as follows:
    \[
    \small \textit{Confidence}\left( \textit{Pattern} \right) = \frac{\textit{Support}\left( \textit{Pattern} \right)}{\textit{Support}\left( ET_1, ET_2 \right)}
    \]
    \item \textit{Lift}: It is the ratio of the confidence of the \textit{Pattern} ($ET_{1}$, $ET_{2}$) $\rightarrow$ $RT_{1}$ to the probability of the relation type $RT_{1}$ appearing independently in the KG, which is used to measure whether there is dependence between ($ET_{1}$, $ET_{2}$) and $RT_{1}$. 
    Its calculation formula is as follows:
    \[
    \small \textit{Lift}\left( \textit{Pattern} \right) = \frac{\textit{Confidence}\left( \textit{Pattern} \right)}{\textit{Support}\left( RT_1 \right)}
    \]
\end{itemize}

In summary, these three metrics evaluate the validity of type triples from different perspectives.
The support reflects the universality, the confidence reflects the reliability of the association, and the lift reflects the interdependence.
However, when the thresholds are set too high, some valid type triples will be missed;
conversely, when the thresholds are set too low, some invalid triples will be retained.
To address this, we design an experiment to explore the optimal choice of thresholds (see Section~\ref{sec: RQ1} for details).
Finally, \textit{Support}, \textit{Confidence}, and \textit{Lift} are set to 0.005, 0.02, and 1.0, respectively.
In this unit, a type triple can be considered valid only when the values of these three metrics of the type triple are all higher than their respective thresholds.
Based on this, we remove the invalid type triples and obtain a validated KG schema.

\subsubsection{KG Update}
With the validated KG schema, we can compare the type information in the KG with the type triples, and then remove the suspicious instance triples in the KG.
Therefore, the inputs to this unit are the validated KG schema and the unreliable KG, and the output is the reliable KG.

\begin{figure*}[t]
    \centering
    \includegraphics[width=1.0\linewidth]{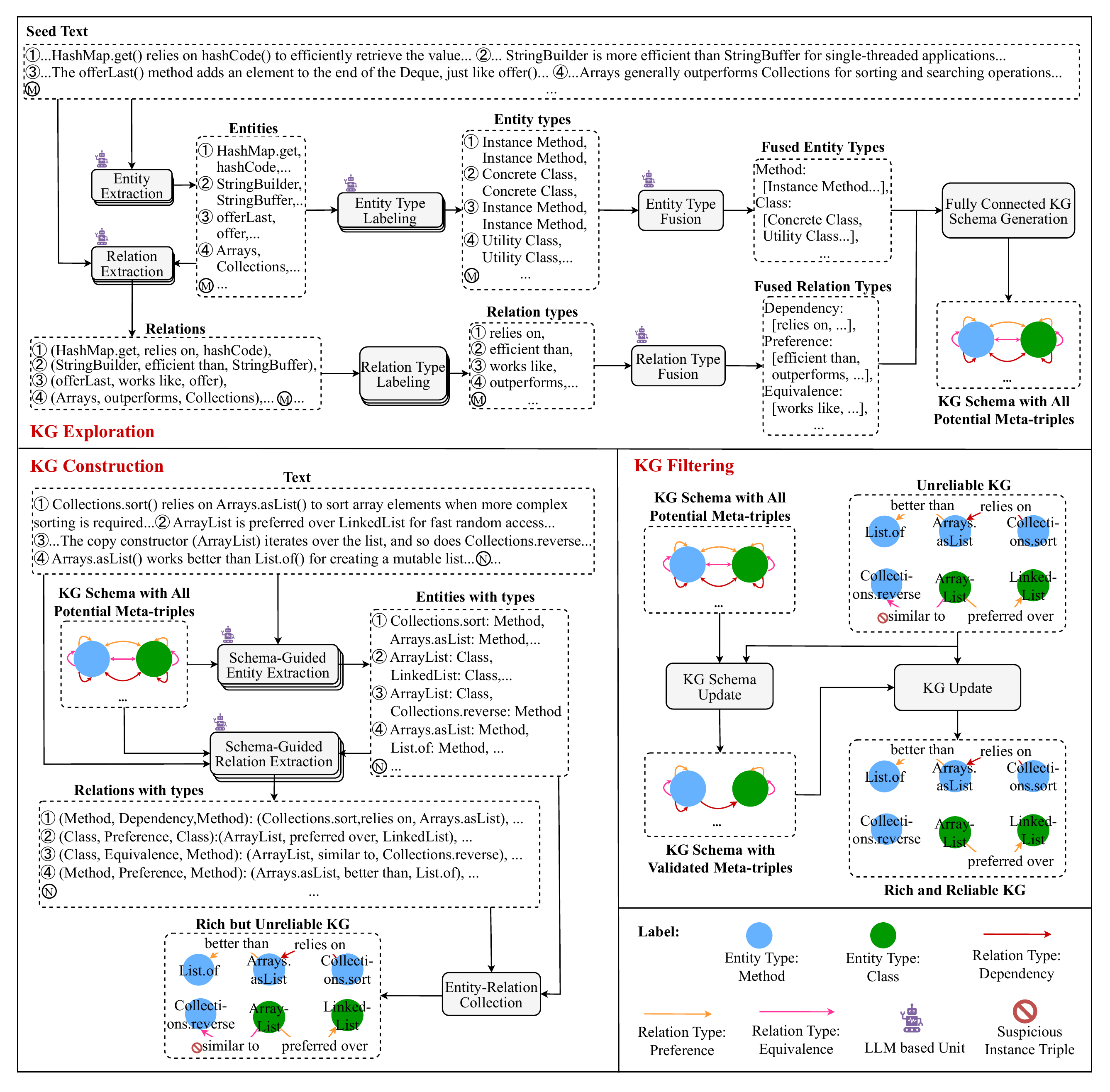}
    \caption{Running Example of Our Framework.}
    \label{fig: running example}
\end{figure*}

\subsection{Running Example}
In this section, we will use an example to explain the execution process of our framework in detail.
As shown in Fig.~\ref{fig: running example}, we select a small portion of text from Stack Overflow posts as the seed text, and another portion as the text to be extracted.
In the KG exploration module, the Seed Texts are first input into the entity extraction unit, which outputs API entities like \textit{Collections} and \textit{Arrays}.
These entities are then combined into API pairs, such as (\textit{Collections}, \textit{Arrays}).
The entity pairs, along with the text, are then input into the relation extraction unit, resulting in instance triples like (\textit{Arrays}, \textit{outperforms}, \textit{Collections}).
Simultaneously, the API entities are sent to the entity type labeling unit to generate the specific entity type for each entity (e.g., utility class), and the instance triples are processed by the relation type labeling unit to obtain the relation type (e.g., outperforms).
The entity type fusion unit then merges all specific entity types into abstract types (e.g., class), while the relation type fusion unit combines all relation types into abstract relation types (e.g., preference). Finally, these entity and relation types are combined to form a comprehensive KG schema, consisting of 12 type triples, e.g., (\textit{class}, \textit{preference}, \textit{class}).
In this schema, as shown in Fig.\ref{fig: running example}, blue and green nodes represent method and class entity types, while red, yellow, and pink edges represent dependency, preference, and equivalence relation types, respectively.

In the KG construction module, all texts and the KG schema containing all potential type triples are input into the schema-guided entity extraction unit, which identifies API entities based on specific entity types (e.g., \textit{Collections.sort}: \textit{static method}).
These entities are then paired to form API pairs.
Next, the API pairs, along with the text and schema, are input into the schema-guided relation extraction unit, which generates instance triples that match specific relation types, e.g., (\textit{method}, \textit{dependency}, \textit{method}): (\textit{Collections.sort}, \textit{relies on}, \textit{ArrayList.asList}).
After all the entities and relations have been extracted from the text, the entity-relation collection unit gathers these instance triples and their type information to construct the API KG, which includes 6 entities and 4 instance triples.
However, because the schema contains invalid type triples (e.g., (\textit{class}, \textit{equivalence}, \textit{method})), some instance triples in the KG are suspicious (e.g., (\textit{ArrayList}, \textit{similar to}, \textit{Collections.reverse})).
To ensure the KG's reliability, these suspicious instance triples must be removed.

As shown in Fig.~\ref{fig: running example}, the KG schema with all potential type triples and the unreliable KG are input into the KG filtering module. 
This module calculates the support, confidence, and lift for each type triple, removing those that do not meet the threshold, along with their corresponding instance triples from the API KG.
For example, the type triple (\textit{class}, \textit{equivalence}, \textit{method}) falls below the threshold, so it is removed from the KG schema, along with the instance triple (\textit{ArrayList}, \textit{similar to}, \textit{Collections.reverse}) from the KG. 
Finally, the KG filtering module outputs a validated KG schema with 6 type triples and a reliable KG containing 6 entities and 3 instance triples.

\section{EXPERIMENTAL SETUP}
This section starts with four research questions for evaluating our method's effectiveness, and then introduces baselines, data preparation, and evaluation metrics.

\subsection{Research Questions (RQs)}
\begin{itemize}[leftmargin=*]
\item What is the optimal threshold in the KG Filtering module?
\item How well does our method perform in KG construction?
\item Is the Explore-Construct-Filter framework effective?
\item How generalizable is our method across different LLMs?
\end{itemize}

\subsection{Baselines}\label{sec: baseline}
In the experiments, we implement a total of seven baseline methods to evaluate our method.
To explore the performance of our method, we propose three baseline methods, namely MKC, APIRI, GraphRAG, and EDC, with details as follows.
\begin{itemize}[leftmargin=*]
    \item Huang et al. \cite{Manual} propose a schema-based method MKC.
    By analyzing the text of API documentation and Stack Overflow posts, they summarized three entity types and nine semantic relation types, and designed a KG schema (see Table~\ref{tab: existing schemas}).
    Finally, they constructed an API KG based on this KG schema using a rule-based method. 
    We obtain its code from Github~\cite{yuangit} and measure the performance of this method by inputting the same test text.
    \item In order to improve the performance of knowledge extraction, Huang et al. propose APIRI~\cite{yanbang2} to extract instance triples based on the KG schema of MKC.
    This method extracts instance triples by querying the LLM for feature knowledge of two APIs (e.g., usage, performance).
    We obtained its code from Github~\cite{APIRICGIT} and tested its performance using the same testing method as MKC.
    \item GraphRAG \cite{GraphRAG} is used to improve the efficiency of retrieving information from the KG.
    It consists of two stages: the indexing stage, where structured data such as entities, relations, and statements are extracted from unstructured text using an LLM, and the querying stage, where information relevant to the user's query is retrieved from the KG.
    This paper focuses only on the first stage, with particular emphasis on entity and relation extraction. 
    Therefore, we obtain the relevant code from Github~\cite{graphraggit} and provide the entity types summarized in MKC to construct an API KG.
    \item Zhang et al.\cite{EDC} design EDC, an automated schema-free method for KG construction. 
    We use their GitHub code\cite{edcgit} to construct an API KG. 
    However, this method normalizes relation instances into relation types but ignores entity types. 
    As a result, the KG schema includes only relation types, and the entity instances lack specific entity types.
\end{itemize}

To evaluate the effectiveness of the Explore-Construct-Filter strategy, we implement three baselines, named Our$_{\text{w/oKE}}$, Our$_{\text{w/oKF}}$, and Our$_{\text{w/oFC}}$, with details as follows.

\begin{itemize}[leftmargin=*]
    \item Our$_{\text{w/oKE}}$ refers to removing the KG exploration module while retaining the KG construction module and the KG filtering module. 
    To ensure the normal execution of our method, we adopted the same KG schema as that of MKC.
    By comparing Our$_{\text{w/oKE}}$ and our method, we can verify whether the KG exploration module can discover more entity types and relation types, thereby enhancing the richness of the KG.
    \item Our$_{\text{w/oKF}}$ means removing the KG filtering module and taking the KG output by the KG construction module as the final result.
    By comparing Our$_{\text{w/oKF}}$ and our method, we can verify whether the KG filtering module can improve the reliability of the API KG.
    \item  Our$_{\text{w/oFC}}$ refers to removing the full combination strategy of entity types and relation types.
    In the KG schema, the type triples are those to which the instance triples in the KG exploration module belong. 
    Compared Our$_{\text{w/oFC}}$ with our method, we can verify if the full combination strategy can generate more reliable type triples to enhance the richness of the API KG, thereby improving the comprehensiveness of the API KG.
\end{itemize}

Note that all methods, including ours and the baselines, are based on the GPT-4o model~\cite{basellm}.
Please refer to the Appendix for the parameter settings of the model.

\subsection{Data Preparation} \label{sec: data pre}
To fairly compare the MKC method~\cite{Manual} with our method, we obtain texts from GitHub that summarize API entity types and relation types for constructing a KG schema. 
These texts are sourced from posts on Stack Overflow and are related to Java APIs, totaling 206 entries.
In this paper, they are treated as seed texts.
Previous work \cite{Manual} also collects 32,505 texts from the Java tutorial~\cite{javatutorial} documentations for constructing the API KG based on KG schema.
However, not all of these texts contain API entities and relations.
For instance, descriptions such as ``This text focuses on the two most common operations: Adding/removing elements...'' do not involve API entities or relations.
To filter out texts that contain API entities and relations, we design three filtering criteria inspired by the API text filtering rules proposed by Huang et al.~\cite{yanbang1}:
\begin{itemize}[leftmargin=*]
    \item Since the method entity usually ends with ``()'', if the text contains ``()'' and the text length is greater than 8 tokens, the text is retained.
    \item The API entity usually contains ``.'' to indicate a function call (e.g., iterator.remove); therefore, if the text contains ``.'' (both before and after are letters) and the text length is greater than 8 tokens, the text is retained.
    \item If the text contains the words ``method'', ``class'', ``package'' (e.g., remove method), and the text length is greater than 8 tokens, the text is retained.
\end{itemize}

\begin{table}[t]
\centering
\caption{The Quality of AI Units}
\label{tab: data_details}
\begin{tabular}{c|c}
\hline
Type Triples                           & Number    \\ \hline
(\textit{class}, \textit{containment}, \textit{method})           &  318      \\ \hline
(\textit{method}, \textit{dependency}, \textit{method})           &  184      \\ \hline
(\textit{class}, \textit{execution}, \textit{method})             &  128      \\ \hline
(\textit{class}, \textit{access}, \textit{method})                &  124      \\ \hline
(\textit{method}, \textit{dependency}, \textit{class})            &  120      \\ \hline
(\textit{method}, \textit{equivalence}, \textit{method})          &  119      \\ \hline
(\textit{method}, \textit{difference}, \textit{method})           &  106      \\ \hline
(\textit{method}, \textit{collaboration}, \textit{method})        &  76       \\ \hline
(\textit{package}, \textit{containment}, \textit{class})          &  75       \\ \hline
(\textit{class}, \textit{implementation}, \textit{interface})     &  72       \\ \hline
\end{tabular}
\end{table}

Using these criteria, we obtain 5,047 texts, which are used to extract API entities and relations for constructing an API KG.
Finally, we discovered 4 entity types and 13 relation types from the seed text, generating 26 verified type triples (see Table~\ref{tab: existing schemas}).
Based on this, we construct a KG containing 1,375 unique entities and 1,843 unique relation instances 
Table~\ref{tab: data_details} shows the ten most frequent type triples in the KG.

However, annotating standard answers for such a large number of texts is extremely labor-intensive.
Therefore, we adopt a sampling method~\cite{Singh1996ElementsOS} that is widely used in previous studies~\cite{Li2018ImprovingAC, Liu2020GeneratingCB}, to ensure that the observed metrics in the sample can be generalized to the entire population.
Thus, at a 95\% confidence level with a confidence interval of 5, we randomly select 384 texts as the test set.
We then invite four PhD students (who are not involved in this study) with over five years of Java development experience to annotate the instance triples in the test set.
During the annotation process, they are divided into two groups, each consisting of two annotators, who independently annotate 192 identical texts.
After the annotation, any conflicts are resolved by an annotator from the other group.
Finally, we calculate the Cohen's Kappa coefficient for the two groups, which are 0.78 and 0.82, indicating almost perfect agreement.
Therefore, we construct the ground truth for this test set, which contains 382 instance triples and 352 unique API entities.

It should be noted that we only annotate instance triples, as they are concrete and can be directly identified from the text, whereas type triples are more abstract, requiring reasoning and categorization, and lack a unified standard, making their annotation more challenging.
To ensure efficient and consistent annotation, we choose to focus solely on instance triples.
The accuracy of type triples can be calculated by annotating the output of the schema-guided entity extraction unit and schema-guided
relation extraction unit, as detailed in Section~\ref{sec: RQ2}.

\subsection{Evaluation Metrics}
In this paper, we employ precision (p), recall (r), and F1-score to evaluate the performance of each method for KG construction.
However, the instance triples extracted by each method may deviate from the ground truth.
For instance, given the sentence ``There is a little difference between forward() and include()...'', the extracted triple could be (\textit{forward()}, \textit{has difference between}, \textit{include()}), whereas the ground truth is (\textit{forward()}, \textit{is different from}, \textit{include()}).
Although the two triples are highly similar, a simple string comparison cannot judge the correctness of the extraction result.

Therefore, we assess the correctness of the extracted triples by calculating their similarity with the ground truth triples.
Specifically, we use a pre-trained BERT model~\cite{devlin2018bert} to generate semantic vectors for the triples and calculate their cosine similarity. It is important to note that we only compute the similarity when both the head entity and the tail entity of the two triples match.
In this study, we define three similarity thresholds: 0.9, 0.92, and 0.94, denoted as @0.9, @0.92, and @0.94, respectively.
Only when the similarity of the extracted triple exceeds the defined threshold is the extraction considered correct.
Based on this, we calculate the precision, recall, and F1 score for instance extraction.




\section{EXPERIMENTAL RESULTS}
This section delves into four RQs to evaluate and discuss our method’s performance.

\subsection{What is the optimal threshold in the KG filtering module?}\label{sec: RQ1}
\subsubsection{Motivation}
In the KG Filtering module, we apply the association rule to filter out invalid type triples.
The association rule involves three key metrics: support, confidence, and lift.
Setting appropriate thresholds for these metrics is crucial for ensuring the effectiveness of the KG filtering module and the reliability of the KG.
This RQ aims to explore the optimal thresholds for these metrics to balance the reliability of the KG with the richness of the API knowledge it contains.

\subsubsection{Methodology}
Due to the extremely large number of possible threshold combinations, it is unrealistic to enumerate all cases. Therefore, we select five representative cases.
In these cases, the support, confidence, and lift are gradually increased from low to high.
The specific details are as follows:
\begin{itemize}[leftmargin=*]
\item Case 1: support = 0.001; confidence = 0.005; lift = 0.6
\item Case 2: support = 0.003; confidence = 0.01;  lift = 0.8
\item Case 3: support = 0.005; confidence = 0.02;  lift = 1.0
\item Case 4: support = 0.007; confidence = 0.03;  lift = 1.2
\item Case 5: support = 0.009; confidence = 0.04;  lift = 1.4
\end{itemize}

We collected the KG schemas generated by the KG filtering module under different thresholds, as well as the KGs constructed based on these schemas.
Then, we collect the outputs corresponding to the texts in the test set, which are used to calculate the experimental metrics P-R-F1.
To assess the effectiveness of the type triples, we invite two PhD students (both with more than 4 years of Java development experience) to annotate the type triples output by the fully connected KG schema generation unit.
To resolve conflicts between them, we assign another PhD student, who does not participate in the annotation process.
Finally, we calculate the Cohen's Kappa coefficient~\cite{Cohen1960ACO}, which results in 0.86, indicating almost perfect agreement among the annotators.
Based on this, we calculate the accuracy of the type triples output by the KG filtering module in each case.

\begin{table*}[t]
\centering
\caption{KG Construction Performance Across Different Thresholds.}
\label{tab: res12}
\begin{tabular}{c|ccc|ccc|ccc|ccl|lll}
\hline
\multirow{2}{*}{Threshold} & \multicolumn{3}{c|}{Case 1}                                  & \multicolumn{3}{c|}{Case 2}                                  & \multicolumn{3}{c|}{Case 3}                                           & \multicolumn{3}{c|}{Case 4}                                                     & \multicolumn{3}{c}{Case 5}                                                     \\ \cline{2-16} 
                           & \multicolumn{1}{c|}{P}    & \multicolumn{1}{c|}{R}    & F1   & \multicolumn{1}{c|}{P}    & \multicolumn{1}{c|}{R}    & F1   & \multicolumn{1}{c|}{P}    & \multicolumn{1}{c|}{R}    & F1            & \multicolumn{1}{c|}{P}    & \multicolumn{1}{c|}{R}    & \multicolumn{1}{c|}{F1} & \multicolumn{1}{c|}{P}    & \multicolumn{1}{c|}{R}    & \multicolumn{1}{c}{F1} \\ \hline
@0.90                      & \multicolumn{1}{c|}{0.53} & \multicolumn{1}{c|}{0.84} & 0.65 & \multicolumn{1}{c|}{0.54} & \multicolumn{1}{c|}{0.84} & 0.66 & \multicolumn{1}{c|}{0.67} & \multicolumn{1}{c|}{0.84} & \textbf{0.75} & \multicolumn{1}{c|}{0.68} & \multicolumn{1}{c|}{0.62} & 0.65                    & \multicolumn{1}{l|}{0.70} & \multicolumn{1}{l|}{0.55} & 0.62                   \\ \hline
@0.92                      & \multicolumn{1}{c|}{0.51} & \multicolumn{1}{c|}{0.82} & 0.63 & \multicolumn{1}{c|}{0.53} & \multicolumn{1}{c|}{0.82} & 0.64 & \multicolumn{1}{c|}{0.66} & \multicolumn{1}{c|}{0.82} & \textbf{0.73} & \multicolumn{1}{c|}{0.68} & \multicolumn{1}{c|}{0.60} & 0.64                    & \multicolumn{1}{l|}{0.69} & \multicolumn{1}{l|}{0.54} & 0.61                   \\ \hline
@0.94                      & \multicolumn{1}{c|}{0.49} & \multicolumn{1}{c|}{0.80} & 0.61 & \multicolumn{1}{c|}{0.51} & \multicolumn{1}{c|}{0.80} & 0.62 & \multicolumn{1}{c|}{0.64} & \multicolumn{1}{c|}{0.80} & \textbf{0.71} & \multicolumn{1}{c|}{0.66} & \multicolumn{1}{c|}{0.59} & 0.62                    & \multicolumn{1}{l|}{0.68} & \multicolumn{1}{l|}{0.53} & 0.60                   \\ \hline
\end{tabular}
\end{table*}

\begin{table}[t]
\centering
\caption{KG Schema Validity Across Different Thresholds.}
\label{tab: res11}
\begin{tabular}{c|c|c|c|c|c}
\hline
Metric & Case 1 & Case 2 & Case 3 & Case 4 & Case 5 \\ \hline
\#Total      & 79              & 48              & 34              & 23               & 27              \\ \hline
\#Correct    & \textbf{30}              & 29              & 26              & 18               & 13              \\ \hline
Accuracy        & 0.43            & 0.60            & 0.76            & \textbf{0.78}            & 0.82           \\ \hline
\end{tabular}
\begin{tablenotes}
\small
\item Note: \#Correct refers to the number of correct type triples among the valid type triples output by the KG filtering module.
\end{tablenotes}
\vspace{-4mm}
\end{table}

\subsubsection{Result}
The validity of the KG schema output by our method in various cases is shown in Table~\ref{tab: res11}.
As the threshold increases, the accuracy of the type triples gradually improves, reaching its highest value of 0.82 in Case 5.
However, the number of correct type triples significantly decreases, with only 14 correct type triples remaining in Case 5.
For our method, an excess of incorrect type triples leads to a large number of suspicious instance triples in the KG, reducing its richness.
On the other hand, too few incorrect type triples result in the loss of some instance triples, impacting the utility of the KG.
Therefore, the KG schema in Case 3 is the most balanced, as its accuracy (0.76) is close to the maximum, while the number of correct type triples (26) remains relatively high.

The KG construction performance of the method is shown in Table~\ref{tab: res12}.
In the three similarity measurement scenarios, as the threshold increases, the method's F1 score first rises and then falls, reaching its highest value in Case 3.
This is because a low threshold introduces incorrect type triples, leading to suspicious instance triples in the KG, which reduces its reliability.
For example, based on incorrect triples (\textit{class}, \textit{equivalence}, \textit{method}), a suspicious instance triple like (\textit{ArrayList}, \textit{similar to}, \textit{Collections.reverse}) can be extracted. 
On the other hand, a high threshold may remove correct type triples and their instance triples, decreasing the richness of the KG.
For example, the KG schema output in Case 5 does not contain the correct type triple (\textit{class}, \textit{difference}, \textit{class}).
Therefore, to balance the reliability and richness of the KG, we adopt the threshold used in Case 3.

\insightbox{
\textbf{Answer:}
Both too low and too high thresholds can impact the effectiveness of the KG.
The threshold in Case 3 is the optimal one, providing the most balanced KG schema.
}

\begin{table*}[t]
\centering
\caption{The Performance of Existing Methods for KG Construction.}
\label{tab: rq22res}
\begin{tabular}{c|ccc|lll|ccc|ccc|ccc}
\hline
\multirow{2}{*}{Threshold} & \multicolumn{3}{c|}{MKC~\cite{Manual}}  & \multicolumn{3}{c|}{APIRI~\cite{yanbang2}} & \multicolumn{3}{c|}{GraphRAG~\cite{GraphRAG}} & \multicolumn{3}{c|}{EDC~\cite{EDC}} & \multicolumn{3}{c}{Our} \\ \cline{2-16} 
& \multicolumn{1}{c|}{P}    & \multicolumn{1}{c|}{R}    & F1   & \multicolumn{1}{c|}{P}    & \multicolumn{1}{c|}{R}    & \multicolumn{1}{c|}{F1} & \multicolumn{1}{c|}{P}    & \multicolumn{1}{c|}{R}    & F1   & \multicolumn{1}{c|}{P}    & \multicolumn{1}{c|}{R}    & F1   & \multicolumn{1}{c|}{P}    & \multicolumn{1}{c|}{R}    & F1   \\ \hline
@0.90                      & \multicolumn{1}{c|}{0.52} & \multicolumn{1}{c|}{0.12} & 0.19 & \multicolumn{1}{l|}{0.51} & \multicolumn{1}{l|}{0.28} & 0.36                    & \multicolumn{1}{c|}{0.31} & \multicolumn{1}{c|}{0.51} & 0.39 & \multicolumn{1}{c|}{0.56} & \multicolumn{1}{c|}{0.62} & 0.59 & \multicolumn{1}{c|}{0.67} & \multicolumn{1}{c|}{0.84} & \textbf{0.75} \\ \hline
@0.92                      & \multicolumn{1}{c|}{0.48} & \multicolumn{1}{c|}{0.11} & 0.18 & \multicolumn{1}{l|}{0.47} & \multicolumn{1}{l|}{0.24} & 0.32                    & \multicolumn{1}{c|}{0.22} & \multicolumn{1}{c|}{0.33} & 0.26 & \multicolumn{1}{c|}{0.54} & \multicolumn{1}{c|}{0.59} & 0.56 & \multicolumn{1}{c|}{0.66} & \multicolumn{1}{c|}{0.82} & \textbf{0.73} \\ \hline
@0.94                      & \multicolumn{1}{c|}{0.37} & \multicolumn{1}{c|}{0.09} & 0.14 & \multicolumn{1}{l|}{0.44} & \multicolumn{1}{l|}{0.20} & 0.28                    & \multicolumn{1}{c|}{0.14} & \multicolumn{1}{c|}{0.18} & 0.15 & \multicolumn{1}{c|}{0.50} & \multicolumn{1}{c|}{0.54} & 0.52 & \multicolumn{1}{c|}{0.64} & \multicolumn{1}{c|}{0.80} & \textbf{0.71} \\ \hline
\end{tabular}
\end{table*}

\begin{table}[t]
\centering
\caption{The Accuracy of AI Units}
\label{tab: rq21res}
\begin{tabular}{c|c}
\hline
AI Unit                           & Accuracy \\ \hline
Entity Extraction                 & 0.83 \\ \hline
Relation Extraction               & 0.78 \\ \hline
Entity Type Labeling              & 0.81 \\ \hline
Entity Type Fusion                & 0.93 \\ \hline
Relation Type Fusion              & 0.82 \\ \hline
Schema-Guided Entity Extraction   & 0.79 \\ \hline
Schema-Guided Relation Extraction & 0.74 \\ \hline
\end{tabular}
\vspace{-4mm}
\end{table}

\subsection{How well does our method perform in KG construction?}\label{sec: RQ2}
\subsubsection{Motivation}
In this paper, we design three main modules to achieve the automated construction of the API KG.
These three modules contain various AI units, such as entity extraction, entity type annotation, and so on.
This RQ aims to explore whether these AI units are effective and to investigate the performance of our method in constructing the API KG.

\subsubsection{Methodology}
We input the seed text into the KG exploration module and collect the outputs (e.g., entities, relations, etc.) of the AI units within it.
Subsequently, we input the test set into the KG construction module and gather the outputs of the AI units within it (e.g., entity types: entity instances, etc.).
Following this, we invite six PhD students (all with over four years of Java development experience) to annotate these results.
The annotation process is as described in Section~\ref{sec: data pre}.
We calculate the Cohen's Kappa coefficient~\cite{Cohen1960ACO} of the annotation results, which were 0.81 and 0.83 respectively, indicating almost perfect agreement among the annotations.
Based on the annotation result, we calculate the accuracy of each unit.
At the same time, we apply all the texts to be extracted to both the existing method and our method, and collect the output results corresponding to the texts in the test set.
We then calculate P-R-F1 score to compare the performance of the methods.

\subsubsection{Result}
The accuracy of the AI units is shown in Table~\ref{tab: rq21res}.
All AI units perform well, with the entity type fusion unit achieving the highest accuracy of 0.93.
It is observed that the more complex the task, the lower the accuracy of the unit.
Additionally, the accuracy of AI units related to relations is generally lower than that of units related to entities.
For example, the schema-guided relation extraction unit has an accuracy of 0.79, as it only needs to ensure the correctness of the identified entities and their types.
On the other hand, the schema-guided relation extraction unit has an accuracy of 0.74, because it not only needs to accurately identify instance triples but also ensure the correctness of their type triples.
Even so, the accuracy of both the schema-guided entity extraction and schema-guided relation extraction units exceeds 70\%, which effectively ensures the reliability of the constructed API KG.

\begin{figure}[t]
    \centering
    \subfloat[]{%
        \includegraphics[width=0.31\linewidth]{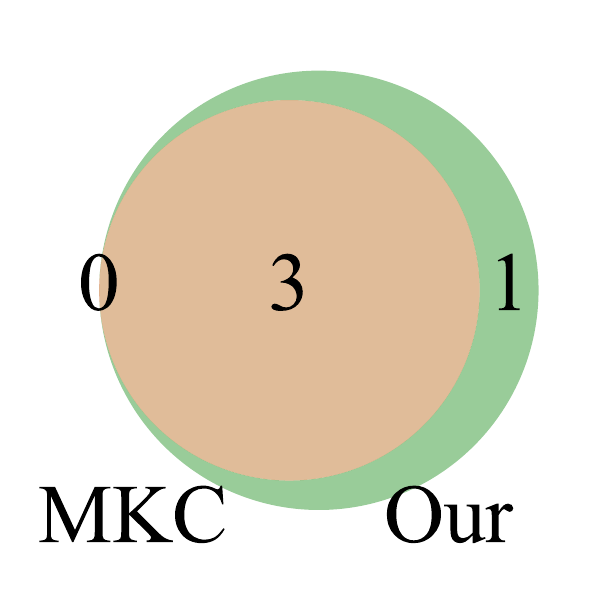}%
    }
    \hfill
    \subfloat[]{%
        \includegraphics[width=0.31\linewidth]{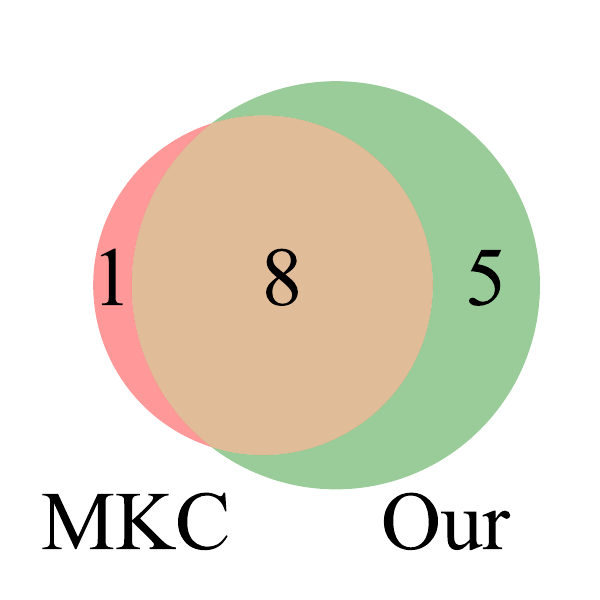}%
    }
    \hfill
    \subfloat[]{%
        \includegraphics[width=0.31\linewidth]{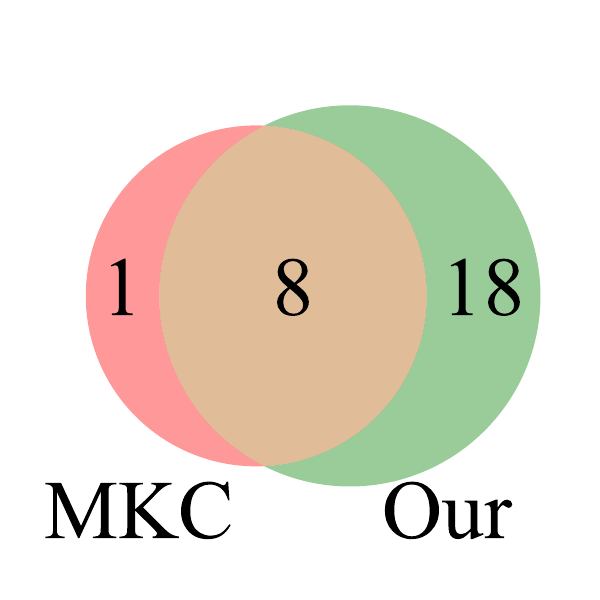}%
    }
    \caption{The Number of Entity Types (a), Relation Types (b) and type triples (c) in Different KG Schemas.}
    \label{fig: schema comparison}
    \vspace{-4mm}
\end{figure}

The comparison of the number of entity types, relation types, and type triples in the KG schemas constructed by the schema-based methods (MKC~\cite{Manual} and our method) is shown in Fig.~\ref{fig: schema comparison}.
For entity types, MKC identifies three entity types, while our method identifies four, with the additional entity type ``interface'' (the specific details are shown in Table~\ref{tab: existing schemas}).
For relation types, MKC and our method share 8 semantically overlapping relation types, but MKC has one unique relation type (function-opposite), while our method has an additional 5 unique relation types (e.g., modification and containment, as detailed in Section~\ref{section: newrel}).
Finally, our method generates 26 correct type triples, of which 8 overlap with MKC's type triples, and the remaining 18 are unique.
MKC also has one unique type triple (\textit{method}, \textit{function opposite}, \textit{method}) due to its unique relation type.
Moreover, although EDC~\cite{EDC} can refine relation types through its strategy, many of these types have similar semantics (e.g., check and test) and could be further merged, as described in Section~\ref{section: newrel}.
In summary, our method can explore a more comprehensive KG schema, laying the foundation for KG construction.

Table~\ref{tab: rq22res} presents the experimental results of different methods for KG construction.
In all three similarity scenarios, our method outperforms existing methods.
Among them, MKC performs the worst.
First, due to the limited entity types and relation types in its KG schema, it is difficult to construct a knowledge-rich KG.
Second, MKC uses a rule-based extraction method, which ensures high precision, but the strict rules result in missing instance triples, leading to low recall and, consequently, poor overall performance.
APIRI~\cite{yanbang2} uses LLM to extract API knowledge, and although its performance improves, it is still limited by the small number of relation types in the KG schema, with a maximum F1 score of only 0.36.
For GraphRAG~\cite{GraphRAG}, the lack of KG schema guidance during the extraction process leads to noise in the results.
For example, GraphRAG incorrectly classifies the Java Virtual Machine (Java VM) as an API entity.
As a result, its highest F1 score is only 0.39.
Additionally, its performance fluctuates significantly across all three scenarios, indicating instability in the prompt design.
GraphRAG uses descriptive sentences to represent relations, failing to accurately capture the textual semantics, which leads to inaccurate relation instances.
The state-of-the-art method EDC performs better than the previous two methods, but its extraction process also lacks KG schema guidance, preventing it from focusing on API entity objects, which leads to noise.
For example, it extracts incorrect instance triples such as (\textit{FilterWriter}, \textit{is a}, \textit{class}).
Compared with EDC, our method improves the F1 score of KG construction by 25.2\%.
Our method extracts API knowledge based on the KG schema with diverse entity types and relation types, enabling the construction of both comprehensive and 
reliable API KG.

\insightbox{
\textbf{Answer:}
The AI units we designed can efficiently complete various 
tasks, ensuring the construction of the KG.
Our method overcomes the limitations of existing method, discovering a rich variety of entity and relation types, thereby constructing a practical and reliable KG.}

\subsection{Is the Explore-Construct-Filter framework effective?}\label{sec: RQ3}
\subsubsection{Motivation}
To enhance the richness and reliability of the API KG, we propose the exploration-construct-filter framework and design three core modules: KG exploration, KG construction, and KG filtering. 
This RQ aims to verify whether this strategy can enhance the effectiveness of the KG.

\subsubsection{Methodology}
In this RQ, we design three variant methods, as described in Section~\ref{sec: baseline}.
For each variant method, we use the same data as in RQ2 (seed texts, texts to be extracted, and the test set) and collect their output results.
By comparing the extracted instance triples with the ground truth in the test set, we calculate precision, recall, and F1 scores to evaluate the performance of each method.

\begin{table*}[t]
\centering
\caption{KG Construction Performance of Variant Methods.}
\label{tab: res32}
\begin{tabular}{c|ccc|ccc|ccc|ccc}
\hline
\multirow{2}{*}{Threshold} & \multicolumn{3}{c|}{Our$_{\text{w/oKE}}$}                    & \multicolumn{3}{c|}{Our$_{\text{w/oKF}}$}                    & \multicolumn{3}{c|}{Our$_{\text{w/oFC}}$}                    & \multicolumn{3}{c}{Our}                                      \\ \cline{2-13} 
                           & \multicolumn{1}{c|}{P}    & \multicolumn{1}{c|}{R}    & F1   & \multicolumn{1}{c|}{P}    & \multicolumn{1}{c|}{R}    & F1   & \multicolumn{1}{c|}{P}    & \multicolumn{1}{c|}{R}    & F1   & \multicolumn{1}{c|}{P}    & \multicolumn{1}{c|}{R}    & F1   \\ \hline
@0.90                      & \multicolumn{1}{c|}{0.59} & \multicolumn{1}{c|}{0.24} & 0.34 & \multicolumn{1}{c|}{0.47} & \multicolumn{1}{c|}{0.84} & 0.60 & \multicolumn{1}{c|}{0.64} & \multicolumn{1}{c|}{0.56} & 0.60 & \multicolumn{1}{c|}{0.67} & \multicolumn{1}{c|}{0.84} & \textbf{0.75} \\ \hline
@0.92                      & \multicolumn{1}{c|}{0.53} & \multicolumn{1}{c|}{0.22} & 0.31 & \multicolumn{1}{c|}{0.44} & \multicolumn{1}{c|}{0.82} & 0.57 & \multicolumn{1}{c|}{0.62} & \multicolumn{1}{c|}{0.55} & 0.58 & \multicolumn{1}{c|}{0.66} & \multicolumn{1}{c|}{0.82} & \textbf{0.73} \\ \hline
@0.94                      & \multicolumn{1}{c|}{0.47} & \multicolumn{1}{c|}{0.21} & 0.29 & \multicolumn{1}{c|}{0.43} & \multicolumn{1}{c|}{0.80} & 0.56 & \multicolumn{1}{c|}{0.60} & \multicolumn{1}{c|}{0.53} & 0.56 & \multicolumn{1}{c|}{0.64} & \multicolumn{1}{c|}{0.80} &\textbf{0.71} \\ \hline
\end{tabular}
\end{table*}

\begin{table}[t]
\centering
\caption{KG Schema Validity of Different Methods.}
\label{tab: res31}
\begin{tabular}{c|c|c|c|c}
\hline
Metric     &  Our$_{\text{w/oKE}}$  & Our$_{\text{w/oKF}}$   & Our$_{\text{w/oFC}}$     & Our \\ \hline
\#Total    &  8                     & 208                    & 20                     & 34   \\ \hline
\#Correct  &  8                     & \textbf{31}                     & 18                     & 26    \\ \hline
Accuracy   &  1.00                  & 0.15                     & \textbf{0.90}                   & 0.76   \\ \hline
\end{tabular}
\vspace{-2mm}
\end{table}

\subsubsection{Result}
The results of the KG schema validity experiment are shown in Table~\ref{tab: res31}.
We can observe that the type triple accuracy of Our$_{\text{w/oKE}}$ is also 1.00 due to it uses the KG schema from MKC.
However, some type triples (e.g., ($method$, $function\ opposite$ $method$)) have a low frequency of occurrence and are removed by the KG filtering module. 
Due to the lack of the KG filtering module, Our$_{\text{w/oKF}}$ retains all 208 type triples generated by the KG exploration module, but only 31 of them are correct, resulting in an accuracy of only 0.15.
On the other hand, Our$_{\text{w/oKF}}$, lacking the fully connected strategy, maintains relatively high accuracy due to the KG filtering module.
However, the number of correct type triples is much smaller than that in our KG schema, thus it cannot comprehensively extract instance triples.
In summary, by comparing these methods, it can be observed that the accuracy of the type triples and the number of correct type triples in our method are relatively high.

As shown in Table~\ref{tab: res32}, the experimental results demonstrate that our method surpasses all variant methods in KG construction.
Specifically, the comparison between Our$_{\text{w/oKE}}$ and our method reveals that the exploration module significantly enhances the richness of the KG, achieving an average improvement of 133.6\% across three similarity thresholds.
The absence of the KG exploration module in Our$_{\text{w/oKE}}$ leads to a lower recall rate for instance triple extraction, primarily because some valid type triples are lost, thereby hindering the extraction of diverse instance triples.
Conversely, Our$_{\text{w/oKF}}$ removes the KG filtering module and includes numerous suspicious instance triples, thereby reducing precision in instance triple extraction.
In contrast, our method incorporates the KG filtering module, which on average improves the richness of the KG by 26.6\%.
Moreover, when compared to Our$_{\text{w/oFC}}$, our method benefits significantly from the fully connected strategy, improving the comprehensiveness of the KG by an average of 33.5\%.
Due to the lack of a fully connected strategy, Our$_{\text{w/oFC}}$ overlooks some valid type triples, which leads to a lower recall rate.
Furthermore, the comparison between Our$_{\text{w/oKE}}$ and Our$_{\text{w/oFC}}$ highlights that the F1 score of Our$_{\text{w/oFC}}$ averages 0.58, significantly higher than the F1 score of Our$_{\text{w/oKE}}$, which further emphasizes the ability of the KG exploration module to uncover a wide array of entity and relation types.

\begin{table}[t]
\centering
\caption{KG Schema Validity across Different LLMs.}
\label{tab: res41}
\begin{tabular}{c|c|c|c}
\hline
Metric       &  Our$_{\text{Llama}}$  & Our$_{\text{Claude}}$        & Our     \\ \hline
\#Total      &  30                     &  32                          & 34      \\ \hline
\#Correct    &  20                    &  23                          & \textbf{26}      \\ \hline
Accuracy     &  0.67                  & 0.72                         & \textbf{0.76}     \\ \hline
\end{tabular}
\end{table}

\insightbox{
\textbf{Answer:}
The explore-construct-filter framework is effective and indispensable.
This framework can significantly improve the richness and reliability of the API KG.

}

\subsection{How generalizable is our method across different LLMs?}\label{sec: RQ5}
\subsubsection{Motivation}
In this paper, we propose an automated method based on the LLM (GPT-4) to construct API KGs.
For this RQ, our goal is to verify whether different LLMs impact our method.

\subsubsection{Methodology}
We select two other popular LLMs, Llama-3.1-70b~\cite{Llama-3.1} and Claude-3.5-Sonnet~\cite{claude-3-5-sonnet}, both of which have demonstrated excellent performance across various natural language processing tasks.
Furthermore, we use the same experimental methodology as before to calculate the metrics for the different methods.

\begin{table*}[t]
\centering
\caption{KG Construction Performance Across Different LLMs.}
\label{tab: rq42}
\begin{tabular}{c|ccc|ccc|ccc}
\hline
\multirow{2}{*}{Threshold} & \multicolumn{3}{c|}{Our$_{\text{Llama}}$}                                   & \multicolumn{3}{c|}{Our$_{\text{Claude}}$}                                  & \multicolumn{3}{c}{Our}                              \\ \cline{2-10} 
                           & \multicolumn{1}{c|}{P}    & \multicolumn{1}{c|}{R}    & F1   & \multicolumn{1}{c|}{P}    & \multicolumn{1}{c|}{R}    & F1   & \multicolumn{1}{c|}{P}    & \multicolumn{1}{c|}{R}    & F1   \\ \hline
@0.90                      & \multicolumn{1}{c|}{0.66} & \multicolumn{1}{c|}{0.63} & 0.65 & \multicolumn{1}{c|}{0.68} & \multicolumn{1}{c|}{0.71} & 0.70 & \multicolumn{1}{c|}{0.67} & \multicolumn{1}{c|}{0.84} & \textbf{0.75} \\ \hline
@0.92                      & \multicolumn{1}{c|}{0.65} & \multicolumn{1}{c|}{0.62} & 0.64 & \multicolumn{1}{c|}{0.67} & \multicolumn{1}{c|}{0.70} & 0.69 & \multicolumn{1}{c|}{0.66} & \multicolumn{1}{c|}{0.82} & \textbf{0.73} \\ \hline
@0.94                      & \multicolumn{1}{c|}{0.64} & \multicolumn{1}{c|}{0.60} & 0.62 & \multicolumn{1}{c|}{0.65} & \multicolumn{1}{c|}{0.67} & 0.66 & \multicolumn{1}{c|}{0.64} & \multicolumn{1}{c|}{0.80} & \textbf{0.71} \\ \hline
\end{tabular}
\end{table*}

\subsubsection{Result}
As shown in Table~\ref{tab: schema3}, both variant methods' KG exploration modules output 4 entity types and 13 relation types, which are identical to the entity and relation types generated by our method.
This indicates that our method demonstrates high stability when exploring KG schema, with minimal influence from the base model.
Furthermore, the accuracy of the type triples output by the KG filtering module for each method is shown in Table~\ref{tab: res41}.
Our method performs the best, with a type triple accuracy of 0.76 and 26 correct type triples.
Our$_{\text{Claude}}$ follows closely, with an accuracy near 0.72, generating 23 correct type triples.
Our$_{\text{Llama}}$ performs slightly weaker, generating 20 correct type triples, but its accuracy still reaches 0.67.
Although the our method outperforms the others in both accuracy and the number of correct type triples, Our$_{\text{Claude}}$ and Our$_{\text{Llama}}$ are still capable of generating relatively accurate type triples to construct a reliable KG.

The performance of KG construction is shown in Table~\ref{tab: rq42}.
Compared to Our$_{\text{Claude}}$ and Our$_{\text{Llama}}$, our method performs better in KG construction.
This is attributed to GPT-4o's training on large-scale corpora and its powerful model parameters, which enhance its information extraction capabilities.
Furthermore, the number and accuracy of the type triples generated by our method outperform those of other methods, improving the performance of KG construction and ensuring that the filtered KG maintains both richness and reliability.
Nevertheless, the highest F1 scores of methods Our$_{\text{Claude}}$ and Our$_{\text{Llama}}$ are still 0.70 and 0.65, respectively, indicating that even when the base model in our method is replaced with a weaker model, it can still generate an effective and reliable KG.
In summary, our method demonstrates good generalizability across different models, and as model performance improves, the method's performance will also be enhanced accordingly.

\insightbox{
\textbf{Answer:}
Our method is universal across different models, and the more capable the model is, the better the performance of this method will be.
}

\section{DISCUSSION}
This section includes two parts: one is the threat to validity of our method, and the other is the advantages of this method.

\vspace{-2mm}
\subsection{Threats to Validity}
The validity of this paper faces three main threats.
The first is the manual annotation of experimental results, which may be influenced by the annotator's subjective judgment. 
To minimize this bias, we assign two annotators to each dataset and calculate the kappa coefficient to assess their agreement. 
All kappa values exceed 0.75, indicating a high level of consistency in the annotation results, thereby ensuring the reliability of the experimental results.

Another threat comes from the threshold setting in the KG filtering module.
If the threshold is set too high, it may filter out some correct and valuable type triples; if set too low, it may retain too many low-confidence type triples, reducing the reliability of the KG.
Since exhaustively testing all threshold combinations is impractical, we test five sets of thresholds and select the optimal one.
However, the chosen threshold may still not be the best, and a few valuable type triples still be filtered out, such as (\textit{method}, \textit{replacement}, \textit{method}).

The last threat comes from the choice of seed text.
To ensure a fair comparison, we treat the text used for designing the KG schema in the baseline~\cite{Manual} as the seed text and construct a KG schema with a more diverse entity types and relation types.
However, we do not test the method's performance on other seed texts.
Nevertheless, we find that the KG schema constructed by our method already covers the majority of type information, including 4 entity types and 13 relation types.
In the future, we will apply this method to more API texts to explore whether additional relation types can be discovered.

\subsection{Advantages of Our Method}
Although this method is currently applied to API data, theoretically, with minor adjustments to the prompts, it could be extended to other data domains, thus becoming a universal method for automatic KG construction.
However, achieving true universality presents some challenges. 
The data structures and semantic characteristics differ significantly across different fields, which may limit the adaptability of this method to other domains. 
In the future, we will focus on adjusting this method to adapt to the specific data structures and semantic requirements of different fields.

Integrating our method with existing KG retrieval technologies, such as GraphRAG~\cite{GraphRAG}, can form a comprehensive tool for knowledge extraction, analysis, and utilization. 
This integration leverages GraphRAG's expertise in KG retrieval and optimization, complementing the intelligent construction capabilities of the proposed method, thereby enabling a one-stop solution for KG construction, updating, and application.
\section{RELATED WORK}
An API KG is a complex network that represents API entities and their relations using a graph structure.
In an API KG, nodes typically represent API entities (e.g., classes, methods, etc.), while edges represent relations between entities (e.g., invocation, constraint, etc.).
The construction of an API KG aims to extract structured knowledge from API documentation, codebases, and other resources, helping developers better understand and utilize APIs.
For example, an API KG can recommend suitable APIs for specific tasks to developers~\cite{huang2023answering}.
Additionally, API KGs can support tasks such as code generation~\cite{liao20243, liu2023codegen4libs} and misuse detection~\cite{ren2023misuse, Ren2020APIMisuseDD}.

Early API KG construction methods~\cite{Ren2021KGAMDAA, Liu2020GeneratingCB, Li2018ImprovingAC} rely on manually designed KG schemas.
These methods typically follow a ``pipeline'' approach, including sub-tasks such as entity identification, entity classification, and relation classification.
For instance, entity identification may use regular expressions~\cite{bacchelli2010linking}, island parsing~\cite{Treude2016AugmentingAD}, or heuristic rules~\cite{Huang2018APIMR}, while relation classification relies on syntactic analysis and annotation techniques~\cite{Liu2020GeneratingCB, li2018improving}.
Additionally, some studies attempt to use machine learning methods to automate the extraction of entities and relations.
For example, Ye et al.~\cite{crf1} propose APIReal, which uses CRF to identify API entities.
Huo et al.~\cite{lstmcrf} design ARCLIN, which uses BI-LSTM as encoder and CRF as decoder to identify API entities, rather than just CRF.
To further improve the accuracy of knowledge extraction, Huang et al. propose AERJE~\cite{yanbang1}, which achieves joint extraction of API entities and relations by fine-tuning a T5 model.
Although these methods perform well in specific scenarios, they also have significant limitations.
On one hand, rule-based methods have limited generalization ability and struggle to adapt to the diversity of different API documents.
On the other hand, machine learning methods typically require large amounts of labeled data, which is particularly challenging in low-resource environments.

With the advancement of LLM technologies, LLM-based KG construction methods have gradually become mainstream.
These methods can be categorized into schema-based and schema-free approaches, depending on whether they rely on predefined KG schemas. Schema-based methods depend on predefined schemas to guide the extraction of entities and relations.
By adding entity and relation types into prompts, they guide the LLM to generate triples that conform to the schema.
For example, Huang et al.~\cite{yanbang2} use GPT-4 to infer API instance triples that align with predefined relation types.
However, schema-based methods also face several challenges: first, schema design requires extensive domain knowledge and manual intervention, which limits automation;
second, manually designed KG schemas may fail to cover all potential relation types, leading to insufficient KG coverage.
For example, Huang et al.~\cite{Manual} only summarize 9 types of semantic relation types, while we discover 13 types.

To reduce manual effort, researchers attempt to build KGs using schema-free methods.
These methods do not rely on predefined schemas and directly extract relation triples from text, making them widely applicable in the field of natural language processing.
For example, Han et al.~\cite{han2023pive} propose the PIVE, which iteratively supplements additional triples by prompting LLMs. Zhang et al.~\cite{EDC} introduce the EDC, which constructs KGs from domain information through steps such as extraction, definition, and normalization.
In our field, researchers focus more on using LLMs to extract API entities rather than relation triples.
For instance, Huang et al. propose PCR~\cite{PCR}, which utilizes Copilot to extract simple names of APIs.
However, schema-free methods also face several challenges.
First, the constructed KGs lack type information, which limits their effectiveness in retrieving specific information.
Second, the relation triples output by LLMs may contain noise.
On the contrary, we propose an automated schema-based API KG construction method.
This method not only alleviates the manual overhead in schema design but also reduces noise through schema-guided API entity-relation extraction.
The KG constructed by this method contains rich type information, significantly enhancing its practicality.

\section{CONCLUSION AND FUTURE WORK}
This paper proposes an LLM-based automated KG construction method to address the issues of high manual cost, and noise in existing method.
The method introduces three key improvements: automation to enhance efficiency, comprehensive type discover to improve richness, and the ``explore-construct-filter'' strategy to ensure reliability.
Specifically, this method includes three core modules: the KG exploration module, the KG construction module, and the KG filtering module.
The KG exploration module generates a complete schema through diverse type combinations based on the ``fully connected graph'' concept, ensuring the schema's completeness and comprehensiveness.
Next, the KG construction module leverages LLMs to extract schema-compliant instances from large-scale text corpora, forming a preliminary KG. Finally, the KG filtering module enhances reliability by filtering out suspicious triples using probabilistic methods.
Experimental results show that this method can explore a schema that includes comprehensive entity types and relation types, and based on this, construct a rich and reliable KG.
While effective with API data, the method's adaptability suggests broader applications across various fields and data structures.
Looking ahead, we aim to integrate this method with KG retrieval tools like GraphRAG to create a comprehensive knowledge extraction, analysis, and utilization toolkit.

\bibliographystyle{unsrt}
\bibliography{sample}

\clearpage

\section*{APPENDIX}\label{sec: appendix}

This section provides supplementary materials, including detailed prompt designs, parameter settings for our method, and the KG schemas generated by different methods.

\subsection{Prompt Design for Our Method}
\begin{figure}[h]  
    \centering
    \includegraphics[width=1.0\linewidth]{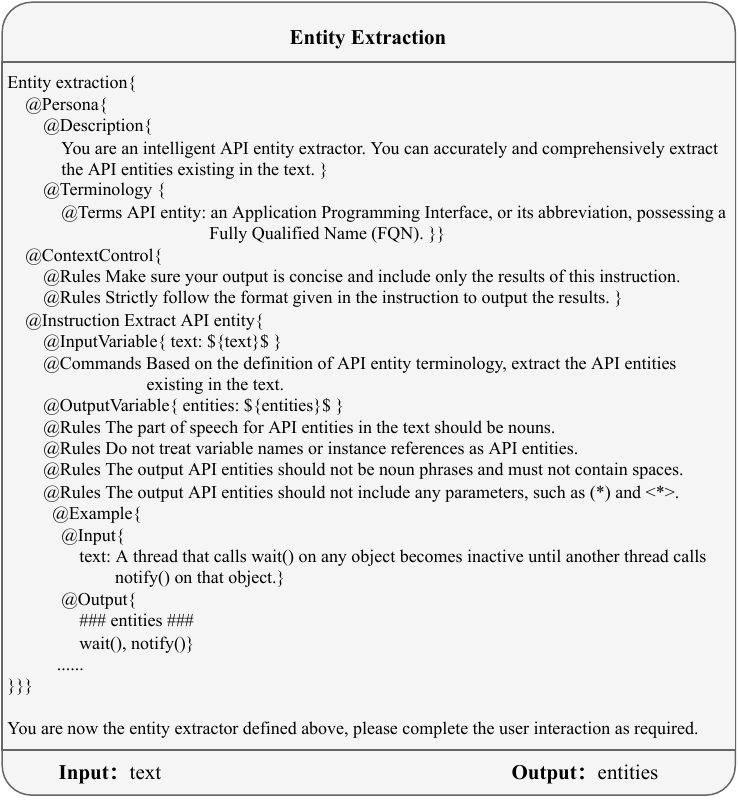}
    \caption{Prompt for Entity Extraction Unit.}
    \label{fig: ee}
\end{figure}

All prompts in this paper use a structured design~\cite{xing2025when}. 
Taking Fig.~\ref{fig: ee} as an example, it has three top-level parts: @Persona (which defines the identity and function of LLM), @ContextControl (which sets behavior constraints for LLM), and @Instruction (which provides operation instructions for LLM).

Among them, Persona contains two sub-parts: 
\begin{itemize}
\item @Description: describes the task objective: (such as ``You are an intelligent API entity extractor...'');
\item @Terminology: describes technical terms: (such as ``Terms API entity...'').
\end{itemize}

@ContextControl contains several @Rules that limit the behavior in the context, e.g., ``Ensure your output is concise...''; 

@Instruction contains five sub-parts:
\begin{itemize}
\item @InputVariable: describes the input of prompt (such as ``text'' here);
\item @Commands: clarifies the execution steps of the LLM, such as ``Based on the definition of API entity terminology, extract the API entities...'';
\item @OutputVariable describes the input of prompt (such as `entities'' here);
\item @Rules: emphasizes the notices when LLM executes the command, such as ``The part of speech for API entities...'', this rule can effectively avoid the common word ambiguity of API entities~\cite{yanbang1}, for example, print'' may be a verb or refer to java.io.printwriter.print();
\item @Example: It is used to help understand the requirements of the task and clarify the output specifications.
\end{itemize}

\clearpage

\clearpage
\subsection{Parameter Setting}\label{sec: paramterset}
In this paper, we implement our method and baselines by calling GPT-4o.
It is the latest model of OpenAI, which has outstanding text understanding capabilities and can perform relatively complex inference tasks~\cite{GPT4oAnalysis-1}.
When calling the LLM, some parameters usually need to be set, including temperature, max\_tokens, n, frequency\_penalty, and presence\_penalty.
Among them, temperature is used to control the randomness of the generated text.
To ensure the stability of our method, we set it to 0 so that the LLM can generate more deterministic results. 
Max\_tokens is used to specify the maximum length of the generated result. 
Since the result lengths output by different units are different, max\_tokens has no fixed value. 
For example, the max\_tokens of the entity extraction unit is set to 128; 
while for the entity type fusion unit and the relation type fusion unit, the max\_token is set to 4096. 
The parameter n represents the number of generated results and is set to 1. 
In addition, frequency\_penalty and presence\_penalty are used to control the coherence of the generated text, and they are kept as the default values (0).

\subsection{The KG Schema Generated by Each Method}\label{section: newrel}
In this section, we will introduce the KG schema generated by each method.
First, Table~\ref{tab: reldetail} presents the 13 relation types generated by our method, 8 of which overlap with the relation types in the existing MKC method~\cite{Manual}.
For example, equivalence and function similarity both indicate that entities are similar or equal in functionality.
Since the occurrence frequency of the relation type``function opposite'' in MKC is low, our method filters it out.
However, we also discover five unique relation types, including: 
\begin{itemize}
\item Containment: It indicates that one entity contains another entity within it.
For example, \textit{SortedMap} contains \textit{headMap()}.
\item Modification: It means that One entity alters or modifies another entity.
For instance, \textit{remove()} can modify the elements in a \textit{SortedSet}.
\item Execution: It represents that one entity initiates or carries out the operation of another entity.
For example, execute \textit{lock()} to close the \textit{Lock} instance.
\item Access: It implies that one entity retrieves or acquires data from another entity.
For example, \textit{readInt()} reads data from a \textit{DataInputStream}.
\item Limitation: It signifies that one entity imposes constraints on another entity’s behavior or
functionality.
For example, the output of \textit{add()} is limited by the state of the \textit{BlockingQueue}.
\end{itemize}
These new relations types form the foundation for constructing a comprehensive API KG.

Table~\ref{tab: existing schemas} compares the differences in type information between existing methods and our method.
GraphRAG~\cite{GraphRAG} and EDC~\cite{EDC} are schema-free methods, with the former lacking relations types (e.g., the type triple (class, null, class)) and the latter lacking entity types (e.g., the type triple (null, check, null)).
In contrast, MKC defines 3 entity types (package, class, method) and 9 relations types, while our method defines 4 entity types (package, class, method, interface) and 13 relation types, resulting in 34 type triples (including 26 correct type triples).
This ensures the comprehensiveness and richness of the KG.
Although the EDC method can refine relation types, there is still redundancy in the final relation types, which can be further optimized.
For example, the relation types such as ``checks'', ``precedes'', and ``test'' have similar semantics and can be further merged.
In contrast, our method can abstract low-dimensional relation types into high-dimensional ones, avoiding such semantic redundancy.

Table~\ref{tab: schema2} shows the KG schemas designed by different variant methods.
Due to Our$_{\text{w/oKE}}$ adopts the schema of MKC, which only contains 3 entity types and 9 relation types, resulting in 9 type triples.
Our$_{\text{w/oKF}}$'s entity and relations types align with ours, but due to the lack of the KG filtering module, it includes 208 type triples, only 31 of which are valid, making the constructed KG unreliable.
Our$_{\text{w/oFC}}$, although consistent with our entity and relation types, lacks a full-connectivity strategy, resulting in only 20 type triples (including 18 correct type triples), making it impossible to construct a comprehensive and rich KG.

Table~\ref{tab: schema3} demonstrates the comparison of the KG schemas designed based on different models.
The results show that while all methods discover the same number of entity and relation types, the knowledge extraction differences lead to discrepancies.
Our$_{\text{Llama}}$ retains 30 type triples, but only 20 of them are correct.
Our$_{\text{Claude}}$ retains 32 type triples, with 23 being correct.
As a result, the KGs constructed by these methods are slightly less rich and reliable compared to the KG constructed by our method.

\clearpage

\begin{table*}[h]
\centering
\caption{The Details of the Relation Types in Our KG Schema.}
\begin{tabularx}{\textwidth}{@{}c|>{\centering\arraybackslash}X|>{\centering\arraybackslash}X@{}}
\hline
Type & Definition & Example \\ \hline
\begin{tabular}[c]{@{}c@{}}Equivalence (function similarity)\end{tabular} & One entity is equal or very similar to another entity in functionality. & The offerLast() method adds an element to the end of the Deque, just like offer(). \\ \hline
\begin{tabular}[c]{@{}c@{}}Difference (behavior difference)\end{tabular} & One entity is different from another, typically in behavior or characteristics. & The add() and offer() methods behave differently when the queue is full. \\ \hline
\begin{tabular}[c]{@{}c@{}}Replacement (function replace)\end{tabular} & One entity can substitute another in certain contexts without changing the expected result. & In many cases, you can replace the File class with the Path interface. \\ \hline
\begin{tabular}[c]{@{}c@{}}Preference (efficiency comparison)\end{tabular} & One entity is favored over another due to efficiency or ease of use in a specific context. & BufferedInputStream is faster than reading single bytes from an InputStream... \\ \hline
\begin{tabular}[c]{@{}c@{}}Dependency (logic constraint)\end{tabular} & One entity depends on another for its functionality or operation. & Collections.sort() relies on Arrays.asList() to sort array elements when more complex sorting is required. \\ \hline
\begin{tabular}[c]{@{}c@{}}Implementation (implement constraint)\end{tabular} & One entity provides a concrete realization or behavior for another entity. & The PoolThreadRunnable class implements the Runnable interface, allowing it to be executed by a thread. \\ \hline
\begin{tabular}[c]{@{}c@{}}Collaboration (function collaboration)\end{tabular} & One entity communicates or works with another entity to complete a specific task. & To set a date on a PreparedStatement or get a date from a ResultSet, you interact with java.sql.Date. \\ \hline
\begin{tabular}[c]{@{}c@{}}Conversion (type conversion)\end{tabular} & One entity is transformed into another entity or format. & You can convert a Set to a List by passing the Set to the addAll() method of a new List. \\ \hline
Containment & One entity contains another entity within it. & The headMap() method of SortedMap returns a new map containing the first elements of the original map. \\ \hline
Modification & One entity alters or modifies another entity. & To remove an element from a SortedSet, you call its remove() method, passing the element to be removed. \\ \hline
Execution & One entity initiates or carries out the operation of another entity. & To lock the Lock instance, you must call its lock() method. \\ \hline
Access & One entity retrieves or acquires data from another entity. & You can read data from a DataInputStream using its readInt() method. \\ \hline
Limitation & One entity imposes constraints on another entity's behavior or functionality. & If the BlockingQueue does not have space for a new element, the add() method throws an IllegalStateException. \\ \hline
\end{tabularx}
\label{tab: reldetail}
\end{table*}

\clearpage

\begin{table*}[t]
\centering
\rotatebox{270}{
\begin{minipage}{\textheight}
  \caption{Comparison of Type Information between the Existing Method and Our Method}
  \label{tab: existing schemas}
  \resizebox{1.0\textheight}{!}{%
    \begin{tabular}{c|cc|cc|cc|cc}
    \hline
    \multirow{2}{*}{Category} &
      \multicolumn{2}{c|}{MKC} &
      \multicolumn{2}{c|}{GraphRAG} &
      \multicolumn{2}{c|}{EDC} &
      \multicolumn{2}{c}{Our} \\ \cline{2-9} 
     &
      \multicolumn{1}{c|}{Number} &
      Content &
      \multicolumn{1}{c|}{Number} &
      Content &
      \multicolumn{1}{c|}{Number} &
      Content &
      \multicolumn{1}{c|}{Number} &
      Content \\ \hline
    Entity Type &
      \multicolumn{1}{c|}{3} &
      package, class, method &
      \multicolumn{1}{c|}{3} &
      package, class, method &
      \multicolumn{1}{c|}{0} &
      - &
      \multicolumn{1}{c|}{4} &
      package, class, method, interface \\ \hline
    Relation Type &
      \multicolumn{1}{c|}{9} &
      \begin{tabular}[c]{@{}c@{}}efficiency comparison,\\ function collaboration,\\ behavior difference,\\ implement constraint,\\ type conversion,\\ logic constraint,\\ function similarity,\\ function opposite, \\ function replace\end{tabular} &
      \multicolumn{1}{c|}{0} &
      - &
      \multicolumn{1}{c|}{53} &
      \begin{tabular}[c]{@{}c@{}}checks,\\ precedes,\\ test,\\ inspects,\\ is called before,\\ created inside,\\ provides,\\ located in package,\\ uses,\\ operates on\\ ...\end{tabular} &
      \multicolumn{1}{c|}{13} &
      \begin{tabular}[c]{@{}c@{}}preference, \\ collaboration,\\ replacement,\\ difference, \\ implementation, \\ conversion, \\ dependency,\\ equivalence,\\ execution, \\ limitation,\\ containment, \\ access,\\ modification\end{tabular} \\ \hline
    Type Triple &
      \multicolumn{1}{c|}{9} &
      \begin{tabular}[c]{@{}c@{}}(class, efficiency comparison, class),\\ (class, function collaboration, class),\\ (package, contain, class),\\ (method, behavior difference, method),\\ (method, implement constraint, method),\\ (class, type conversion, class),\\ (method, logic constraint, method),\\ (method, function similarity, method),\\ (class, has method, method),\\ (method, function opposite, method),\\ (method, function replace, method)\end{tabular} &
      \multicolumn{1}{c|}{9} &
      \begin{tabular}[c]{@{}c@{}}(class, null, class),\\ (class, null, method),\\ (class, null, package),\\ (method, null, method),\\ (method, null, class),\\ (method, null, package),\\ (package, null, package),\\ (package, null, class),\\ (package, null, method)\end{tabular} &
      \multicolumn{1}{c|}{53} &
      \begin{tabular}[c]{@{}c@{}}(null, check, null),\\ (null, precedes, null),\\ (null, test, null),\\ (null, inspects, null),\\ (null, is called before, null),\\ (null, created inside, null),\\ (null, provides, null),\\ (null, located in package, null),\\ (null, uses, null),\\ ...\end{tabular} &
      \multicolumn{1}{c|}{34} &
      \begin{tabular}[c]{@{}c@{}}(class, preference, class),\\ (class, collaboration, class),\\ (package, containment, class),\\ (method, difference, method),\\ (class, implementation, class),\\ (class, conversion, class),\\ (method, dependency, interface),\\ (class, equivalence, class),\\ (method, execution, method),\\ (method, limitation, method),\\ (method, replacement, method),\\ (class, access, method),\\ (method, modification, class), \\ (class, difference, class),\\ (method, preference, method), \\ ...\end{tabular} \\ \hline
    \end{tabular}
  }
\end{minipage}
}
\end{table*}

\begin{table*}[t]
\centering
\rotatebox{270}{
\begin{minipage}{\textheight}
  \caption{Comparison of KG Schemas between Variant Methods}
  \label{tab: schema2}
  \resizebox{1.0\textheight}{!}{%
    \begin{tabular}{ccccccccc}
    \hline
    \multirow{2}{*}{Category} &
      \multicolumn{2}{c}{Our$_{\text{w/oKE}}$} &
      \multicolumn{2}{c}{Our$_{\text{w/oKF}}$} &
      \multicolumn{2}{c}{Our$_{\text{w/oFC}}$} &
      \multicolumn{2}{c}{Our} \\ \cline{2-9} 
     &
      \multicolumn{1}{c|}{Number} &
      \multicolumn{1}{c|}{Content} &
      \multicolumn{1}{c|}{Number} &
      \multicolumn{1}{c|}{Content} &
      \multicolumn{1}{c|}{Number} &
      \multicolumn{1}{c|}{Content} &
      \multicolumn{1}{c|}{Number} &
      Content \\ \hline
    \multicolumn{1}{c|}{Entity Type} &
      \multicolumn{1}{c|}{3} &
      \multicolumn{1}{c|}{package, class, method} &
      \multicolumn{1}{c|}{4} &
      \multicolumn{1}{c|}{package, class, method, interface} &
      \multicolumn{1}{c|}{4} &
      \multicolumn{1}{c|}{package, class, method, interface} &
      \multicolumn{1}{c|}{4} &
      package, class, method, interface \\ \hline
    \multicolumn{1}{c|}{Relation Type} &
      \multicolumn{1}{c|}{8} &
      \multicolumn{1}{c|}{\begin{tabular}[c]{@{}c@{}}efficiency comparison,\\ function collaboration,\\ behavior difference,\\ implement constraint,\\ type conversion,\\ logic constraint,\\ function similarity,\\ function replace, \end{tabular}} &
      \multicolumn{1}{c|}{13} &
      \multicolumn{1}{c|}{\begin{tabular}[c]{@{}c@{}}preference, \\ collaboration,\\ replacement,\\ difference, \\ implementation, \\ conversion, \\ dependency,\\ equivalence,\\ execution, \\ limitation,\\ containment, \\ access,\\ modification\end{tabular}} &
      \multicolumn{1}{c|}{13} &
      \multicolumn{1}{c|}{\begin{tabular}[c]{@{}c@{}}preference, \\ collaboration,\\ containment,\\ difference, \\ implementation, \\ conversion, \\ dependency,\\ equivalence,\\ execution, \\ limitation,\\ replacement,\\  access,\\ modification\end{tabular}} &
      \multicolumn{1}{c|}{13} &
      \begin{tabular}[c]{@{}c@{}}preference, \\ collaboration,\\ containment,\\ difference, \\ implementation, \\ conversion, \\ dependency,\\ equivalence,\\ execution, \\ limitation,\\ replacement, \\ modification,\\ access\end{tabular} \\ \hline
    \multicolumn{1}{c|}{Type Triple} &
      \multicolumn{1}{c|}{8} &
      \multicolumn{1}{c|}{\begin{tabular}[c]{@{}c@{}}(class, efficiency comparison, class),\\ (class, function collaboration, class),\\ (package, contain, class),\\ (method, behavior difference, method),\\ (method, implement constraint, method),\\ (class, type conversion, class),\\ (method, logic constraint, method),\\ (method, function similarity, method),\\ (method, function replace, method)\end{tabular}} &
      \multicolumn{1}{c|}{208} &
      \multicolumn{1}{c|}{\begin{tabular}[c]{@{}c@{}}(class, preference, class),\\ (class, collaboration, class),\\ (package, containment, class),\\ (method, difference, method),\\ (class, implementation, class),\\ (class, conversion, class),\\ (method, dependency, interface),\\ (class, equivalence, class),\\ (method, execution, method),\\ (method, limitation, method),\\ (method, replacement, method),\\ (class, access, method),\\ (method, modification, class), \\ (class, difference, class),\\ (method, preference, method), \\ ...\end{tabular}} &
      \multicolumn{1}{c|}{20} &
      \multicolumn{1}{c|}{\begin{tabular}[c]{@{}c@{}}(class, preference, class),\\ (method, collaboration, method),\\ (package, containment, method),\\ (method, difference, method),\\ (class, implementation, class),\\ (class, conversion, class),\\ (method, dependency, method),\\ (method, equivalence, method),\\ (method, execution, method),\\ (method, limitation, method),\\ (method, replacement, method),\\ (method, modification, class), \\ ...\end{tabular}} &
      \multicolumn{1}{c|}{34} &
      \begin{tabular}[c]{@{}c@{}}(class, preference, class),\\ (class, collaboration, class),\\ (package, containment, class),\\ (method, difference, method),\\ (class, implementation, class),\\ (class, conversion, class),\\ (method, dependency, interface),\\ (class, equivalence, class),\\ (method, execution, method),\\ (method, limitation, method),\\ (method, replacement, method),\\ (class, access, method),\\ (method, modification, class), \\ (class, difference, class),\\ (method, preference, method), \\ ...\end{tabular} \\ \hline
\end{tabular}
  }
\end{minipage}
}
\end{table*}

\begin{table*}[t]
\centering
\rotatebox{270}{
\begin{minipage}{\textheight}
  \caption{Comparison of KG Schemas across Different Models}
  \label{tab: schema3}
  \resizebox{1.0\textheight}{!}{%
    \begin{tabular}{c|cc|cc|cc}
    \hline
    \multirow{2}{*}{Category} &
      \multicolumn{2}{c|}{Our$_{\text{Llama}}$} &
      \multicolumn{2}{c|}{Our$_{\text{Claude}}$} &
      \multicolumn{2}{c}{Our} \\ \cline{2-7} 
     &
      \multicolumn{1}{c|}{Number} &
      Content &
      \multicolumn{1}{c|}{Number} &
      Content &
      \multicolumn{1}{c|}{Number} &
      Content \\ \hline
    Entity Type &
      \multicolumn{1}{c|}{4} &
      package, class, method, interface &
      \multicolumn{1}{c|}{4} &
      package, class, method, interface &
      \multicolumn{1}{c|}{4} &
      package, class, method, interface \\ \hline
    Relation Type &
      \multicolumn{1}{c|}{13} &
      \begin{tabular}[c]{@{}c@{}}preference, \\ collaboration,\\ containment,\\ difference, \\ implementation, \\ conversion, \\ dependency,\\ equivalence,\\ execution, \\ limitation,\\ replacement, \\ access,\\ modification\end{tabular} &
      \multicolumn{1}{c|}{13} &
      \begin{tabular}[c]{@{}c@{}}preference, \\ collaboration,\\ containment,\\ difference, \\ implementation, \\ conversion, \\ dependency,\\ equivalence,\\ execution, \\ limitation,\\ replacement, \\ access,\\ modification\end{tabular} &
      \multicolumn{1}{c|}{13} &
      \begin{tabular}[c]{@{}c@{}}preference, \\ collaboration,\\ containment,\\ difference, \\ implementation, \\ conversion, \\ dependency,\\ equivalence,\\ execution, \\ limitation,\\ replacement, \\ modification,\\ access\end{tabular} \\ \hline
    Type Triple &
      \multicolumn{1}{c|}{30} &
      \begin{tabular}[c]{@{}c@{}}(class, preference, class),\\ (class, collaboration, class),\\ (package, containment, class),\\ (method, difference, method),\\ (class, implementation, class),\\ (class, conversion, class),\\ (method, dependency, interface),\\ (class, equivalence, class),\\ (method, execution, method),\\ (method, limitation, method),\\ (method, replacement, method),\\ (class, access, method),\\ (method, modification, class), \\ (class, difference, class),\\ (method, preference, method), \\ ...\end{tabular} &
      \multicolumn{1}{c|}{32} &
      \begin{tabular}[c]{@{}c@{}}(class, preference, class),\\ (class, collaboration, class),\\ (package, containment, class),\\ (method, difference, method),\\ (class, implementation, class),\\ (class, conversion, class),\\ (method, dependency, interface),\\ (class, equivalence, class),\\ (method, execution, method),\\ (method, limitation, method),\\ (method, replacement, method),\\ (class, access, method),\\ (method, modification, class), \\ (class, difference, class),\\ (method, preference, method), \\ ...\end{tabular} &
      \multicolumn{1}{c|}{34} &
      \begin{tabular}[c]{@{}c@{}}(class, preference, class),\\ (class, collaboration, class),\\ (package, containment, class),\\ (method, difference, method),\\ (class, implementation, class),\\ (class, conversion, class),\\ (method, dependency, interface),\\ (class, equivalence, class),\\ (method, execution, method),\\ (method, limitation, method),\\ (method, replacement, method),\\ (class, access, method),\\ (method, modification, class), \\ (class, difference, class),\\ (method, preference, method), \\ ...\end{tabular} \\ \hline
\end{tabular}
  }
\end{minipage}
}
\end{table*}

\end{sloppypar}
\end{document}